\documentclass[%
 reprint,
superscriptaddress,
 amsmath,amssymb,
 aps,
floatfix,
]{revtex4-2}

\usepackage{amsthm}
\newtheorem{theorem}{Theorem}
\newtheorem{lemma}{Lemma}
\newtheorem{definition}{Definition}

\usepackage{siunitx}
\usepackage{graphicx}
\usepackage{dcolumn}
\usepackage{bm}
\usepackage[colorlinks,linkcolor=blue, citecolor=red]{hyperref}

\begin{document}

\preprint{APS/123-QED}

\title{Molecular Entanglement Witness by Absorption Spectroscopy in Cavity QED}

\author{Weijun Wu}
    \affiliation{Department of Chemistry, Princeton University, Princeton, New Jersey, 08544, USA}

\author{Francesca Fassioli}
    \affiliation{Scuola Internazionale Superiore di Studi Avanzati, Trieste 34136, Italy}

\author{David A. Huse}
    \affiliation{Department of Physics, Princeton University, Princeton, New Jersey, 08544, USA}

\author{Gregory D. Scholes}
    \email{gscholes@princeton.edu}
    \affiliation{Department of Chemistry, Princeton University, Princeton, New Jersey, 08544, USA}

\date{\today}

\begin{abstract}
Producing and maintaining molecular entanglement at room temperature and detecting multipartite entanglement features of macroscopic molecular systems remain key challenges for understanding inter-molecular quantum effects in chemistry. Here, we study the quantum Fisher information, a central concept in quantum metrology, as a multipartite entanglement witness. We generalize the entanglement witness functional related to quantum Fisher information regarding non-identical local response operators. We show that it is a good inter-molecular entanglement witness for ultrastrong light-matter coupling in cavity quantum electrodynamics, including near the superradiant phase transition. We further connect quantum Fisher information to the dipole correlator, which suggests that this entanglement could be detected by absorption spectroscopy. Our work proposes a general protocol to detect inter-molecular entanglement in chemical systems at room temperature.
\end{abstract}

\maketitle


\textit{Introduction.}—Quantum entanglement, a special form of non-classical correlations, is considered a fundamental concept of many-body physics \cite{amico2008entanglement, horodecki2009quantum, abanin2019colloquium} and an important resource in quantum information \cite{weedbrook2012gaussian, mcardle2020quantum, ladd2010quantum}. Inter-atomic entanglement can be revealed and manipulated in many ultra-cold atom systems \cite{saffman2010quantum}. However, studying inter-molecular entanglement in chemical systems is a challenge, because the existence of long-lived entanglement, the measurement of multipartite entanglement, and the non-trivial quantum role that entanglement plays are still not fully understood or demonstrated \cite{wu2024foundations}. 

Inter-molecular entanglement is produced by the interaction between molecules.
However, system-bath interaction induces information spreading into the bath \cite{nandkishore2015many} that causes the system to dephase and lose entanglement. Different from well-isolated ultra-cold atoms, problems of interest in chemistry often involve room-temperature molecular systems in which entanglement is short-lived \cite{wang2021single} and elusive because the inter-molecular interaction is overwhelmed by couplings to the environment.

\begin{figure}[htbp]
    \centering
    \includegraphics[width=0.48\textwidth]{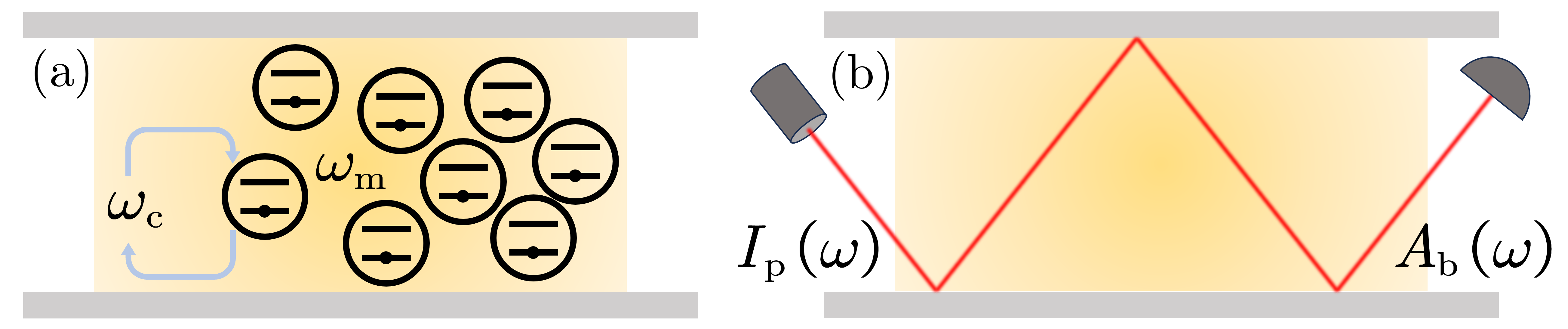}
    \caption{(a) Cavity QED setup, where the $N_{\mathrm{B}}$ molecules of excitation energy $\omega_{\mathrm{m}}$ are collectively coupled to the monochromatic cavity mode $\omega_{\mathrm{c}}$. (b) Bound mode absorption spectrum $A_{\mathrm{b}}\left( \omega \right)$, for pumping laser driving $I_{\mathrm{p}}\left( \omega \right)$ with in-plane wave-vector beyond the threshold of total internal reflection \cite{herrera2017absorption}.}
    \label{fig:experiments}
\end{figure}

To build and maintain entanglement in the presence of this environment-induced decoherence, stronger inter-molecular coupling is needed. The recent development of cavity quantum electrodynamics (cavity QED) \cite{frisk2019ultrastrong,raimond2001manipulating,garcia2021manipulating} in chemistry by embedding a layer of molecules into a pair of mirrors (Fig.~\ref{fig:experiments}(a)) can achieve strong light-matter interaction \cite{forn2019ultrastrong} and create the novel hybridized light-dressed molecular quantum states known as polaritons \cite{mandal2022theoretical}
. Because the cavity mode can couple to many molecules, cavity QED induces an enhancement of light-matter coupling known as collective coupling $G=\sqrt{N_{\mathrm{B}}}g$, where $g$ is the single molecule-cavity interaction and $N_{\mathrm{B}}$ is the number of molecules coherently coupled to the cavity.  The collective nature of molecular polaritons is central to delocalization phenomena including long-range exciton migration  \cite{zhong2017energy, rozenman2018long} and charge transfer \cite{hagenmuller2017cavity, semenov2019electron, wu2022polariton} in organic materials, where spatially distant molecules can be correlated. This implies that cavity QED may have the potential to induce and maintain long-lived multipartite entanglement among molecules\cite{demirchyan2014qubits}.

To detect multipartite entanglement features for a macroscopic-sized many-body system (e.g. the chemical system) is a challenge. These systems are not accessible to quantum tomography \cite{jurcevic2014quasiparticle, lvovsky2009continuous}, to direct measurement of the entanglement entropy
, or the Peres–Horodecki criterion \cite{horodecki2001separability}, due to the limitation of exponential scaling complexity. Despite NP-hardness of the sufficient and necessary conditions for entanglement \cite{gurvits2004classical}, there are some sufficient but not necessary conditions for entanglement, quantified by entanglement witnesses (EW) \cite{friis2019entanglement, guhne2009entanglement, terhal2000bell}. An EW is a function $\mathcal{W} \left( \rho \right)$, whose negative value $\mathcal{W} \left( \rho \right) <0 $ means that quantum state $\rho$ is entangled, while non-negative values are consistent with both entangled and non-entangled states. The choice of EW is not unique and its utility varies based on the system and states of interest. In this paper, we focus on an EW based on quantum Fisher information (QFI).

Generalized from classical Fisher information
, quantum Fisher information $F_{\mathrm{Q}}\left[ \rho ,O \right]$ was originally studied to evaluate the sensitivity of state $\rho$ under certain perturbation $O$. The phase estimation error of a unitary transformation $\rho \left( \theta \right) =e^{-i\theta O}\rho e^{i\theta O}$ is given by the quantum Cramér–Rao bound \cite{braunstein1994statistical}: $\left( \Delta \theta _{\mathrm{est}} \right) ^2\geqslant \frac{1}{F_{\mathrm{Q}}\left[ \rho ,O \right]}$. QFI for any given $\rho$ (diagonalized as $\rho = \sum_l p_l |l\rangle\langle l|$) and $O$ can be expressed as
\begin{equation}
F_{\mathrm{Q}}\left[ \rho ,O \right] = 4\sum_{l,{l^{\prime}}}^{\mathrm{dim}\left( \rho \right)}{p_l\frac{p_l-p_{l^{\prime}}}{p_l+p_{l^{\prime}}}\left| \langle l|O|{l^{\prime}}\rangle \right|^2}~.
\label{eq:QFI1}
\end{equation}
Other than the application in quantum-enhanced sensing and quantum metrology \cite{liu2020quantum, giovannetti2006quantum}, QFI has been proposed as a signature for quantum phase transitions \cite{wang2014quantum, gu2010fidelity} and spin squeezing \cite{ma2009fisher}. Previous work also showed that a system of $N$ identical particles must be multipartite entangled if its QFI regarding $N$ identical local response operators exceeds certain thresholds \cite{hyllus2012fisher, sifain2021toward, hauke2016measuring}. However, in many-body systems, the local operators are usually not identical to each other. For example, in cavity QED, the field operators and matter operators are different; the coupling operator to the cavity field generally varies between molecules. Thus, a universal EW related to QFI regarding non-identical local operators is of interest.

In this paper, we generalize the QFI inequality for non-identical local response operators and further propose an experimentally feasible protocol for multipartite entanglement witness based on the generalized QFI inequality. We test its validity and efficiency for thermal states and polariton states in the ultrastrong coupling regime described by the squeezed Dicke model. We numerically study the simplest multipartite entanglement case of three molecules $\left(N_\mathrm{B}=3\right)$, followed by the analytic expression of QFI in the thermodynamic limit $\left(N_{\mathrm{B}}\rightarrow \infty \right)$ and the singularity during the superradiant phase transition, to demonstrate the long-lived multipartite entanglement in cavity QED. Finally, we show that measuring QFI at thermodynamic equilibrium or pure states is experimentally feasible, in particular, via bounded mode absorption for cavity QED.

\textit{Generalized QFI inequality.}---Consider a system consisting of $N$ different ``particles'', each perturbed by local operators $\left\{ O_i \right\} $. The total response operator is $O=\sum_{i=1}^N{O_i}$. Each local operator has its spectrum $O_i|\lambda _{l,i}\rangle =\lambda _{l,i}|\lambda _{l,i}\rangle$. Without loss of generality, relabel the particles in descending order of local operator spectrum width, $\Delta _1\geqslant \Delta _2\geqslant \cdots \geqslant \Delta _N$, where $\Delta _i=\underset{l}{\max}\left\{ \lambda _{l,i} \right\} -\underset{l}{\min}\left\{ \lambda _{l,i} \right\} $. If $N$-particle state $\rho$ is of entanglement depth $K$, QFI is upper bounded by $F_{\mathrm{Q}}\left[ \rho ,O \right] \leqslant F\left( K;\left\{ \Delta _i \right\} \right)$, where
\begin{equation}
F\left( K;\left\{ \Delta _i \right\} \right) =\sum_{l=0}^{q-1}{\left( \sum_{j=1}^K{\Delta _{Kl+j}} \right) ^2+\left( \sum_{j=1}^r{\Delta _{Kq+j}} \right) ^2}
 ~.
\label{eq:QFIupperbound}
\end{equation}
Here, $q=\lfloor N/K \rfloor$ and $r=N-Kq$ are the quotient and remainder. See Supplemental Material  \cite{SupplementalMaterial} for the rigorous proof. The intuition of Eq.~(\ref{eq:QFIupperbound}) is a state consisting of the first $K$ particles maximally entangled (GHZ state), the next $K$ particles maximally entangled, ..., and the last $r$ particles maximally entangled. $F\left( K;\left\{ \Delta _i \right\} \right)$ is a monotonic increasing function with respect to $K$. 

This suggests a multipartite entanglement witness:
\begin{equation}
\mathcal{W} _{\mathrm{QFI}}\left( \rho ;k \right)  =F\left( k;\left\{ \Delta _i \right\} \right) -F_{\mathrm{Q}}\left[ \rho ,O \right] ~.
\label{eq:EWfunctional}
\end{equation}
The inequality $F_{\mathrm{Q}}\left[ \rho ,O \right] >F\left( k;\left\{ \Delta _i \right\} \right)$ means that the entanglement depth \cite{friis2019entanglement} of $\rho$ must be in the range $N\geqslant K\geqslant k+1$. In particular, $F_{\mathrm{Q}}\left[ \rho ,O \right] >F\left( 1;\left\{ \Delta _i \right\} \right)$ shows that $\rho$ is entangled; $F_{\mathrm{Q}}\left[ \rho ,O \right] >F\left( N-1;\left\{ \Delta _i \right\} \right) $ demonstrates genuine multipartite entanglement (GME \cite{friis2019entanglement, amico2008entanglement}, i.e. fully inseparable, $K=N$). Thus, $F_{\mathrm{Q}}\left[ \rho ,O \right]$ indicates a lower bound of entanglement depth. 

One may consider another case where the total system is partially perturbed, i.e. a system $\rho$ containing $N$ different particles, $N_{\mathrm{B}}$ of which, known as subsystem B, are each perturbed by local operators $\left\{ O_i \right\} $ while the rest $N_{\mathrm{A}} = N - N_{\mathrm{B}}$ particles (subsystem A) are left unperturbed. This is a special case for the generalized QFI inequality where $\Delta _{i>N_{\mathrm{B}}}=0$ and $O=O_{\mathrm{B}}=\sum_{i=1}^{N_{\mathrm{B}}}{O_i}$. Eq.~(\ref{eq:QFIupperbound}) remains the same but $q$ and $r$ are replaced by $q_{\mathrm{B}}=\lfloor N_{\mathrm{B}} / K \rfloor$ and $r_{\mathrm{B}}=N_{\mathrm{B}}-K{q_{\mathrm{B}}}$ respectively. Since $O$ does not couple to subsystem A, only the entanglement among subsystem B contributes to the witness Eq.~(\ref{eq:EWfunctional}). Thus, QFI witnesses not only the entanglement depth but also the pattern of multipartite entanglement: $F_{\mathrm{Q}}\left[ \rho ,O_{\mathrm{B}} \right] >F\left( k;\left\{ \Delta _i \right\} \right)$ means that for any convex factorization $\rho =\sum_{\nu}{p_{\nu}\left( \bigotimes_l{\rho _{\nu ,l}} \right)}$, there must exist certain $\rho _{\nu ,l}$ containing $k+1$ or more particles from subsystem B. See Supplemental Material \cite{SupplementalMaterial} for the rigorous proof.

\textit{Measuring QFI.}—An advantage of QFI 
as an EW is that QFI can be derived from other observables. Thus Eq.~(\ref{eq:EWfunctional}) is an experimentally feasible protocol. Eq.~(\ref{eq:QFI1}) shows that QFI depends on the eigenstructure of $\rho$, so QFI for a general state $\rho$ is not observable. Luckily, for some special cases of experimental interest, namely, pure states and also thermal states, QFI is measurable.

The QFI of a pure state can be reduced to the variance:
$F_{\mathrm{Q}}\left[ |\psi \rangle ,O \right] =4\left( \langle \psi |O^2|\psi \rangle -\langle \psi |O|\psi \rangle ^2 \right) $. Only the expectation values of $O$ and $O^2$ need to be measured. 

The QFI of a thermal state $\rho \left( T \right) =e^{-H/T}/Z$ can be written in the Kubo response framework: 
\begin{equation}
F_{\mathrm{Q}}\left[ \rho \left( T \right) ,O \right] =4\int_0^{\infty}{d\omega \frac{\left( 1-e^{-\omega /T} \right) ^2}{1+e^{-\omega /T}}I\left( \omega ;T \right)}~,
\label{eq:QFIfromSpectrum}
\end{equation}
where $I\left( \omega  \right)$ is the time correlation function of $\tilde{O}\left( t \right) =e^{iHt}Oe^{-iHt}$ in the frequency domain:
\begin{equation}
    \begin{aligned}
        I\left( \omega \right) 
        =&
        \frac{1}{\pi}\mathrm{Re}\left[ \int_0^{\infty}{dte^{i\omega t}\left< \tilde{O}\left( t \right) \tilde{O}\left( 0 \right) \right>} \right]
        \\
        =&
        \sum_{l,l^{\prime}}{{p_l}\left| \langle l|O|l^{\prime}\rangle \right|^2\delta \left( \omega -\omega _{l^{\prime}}+\omega _l \right)}~,
    \end{aligned}
    \label{eq:Spectrumfromcorrelation}
\end{equation}
where $p_l$ is the Boltzmann distribution of eigenstate $|l\rangle$. The intuition of Eq.~(\ref{eq:QFIfromSpectrum}) is that $I\left( \omega \right)$ measures the sensitivity of response to the perturbation. A similar expression shows the relation between thermal QFI and dynamic susceptibility \cite{hauke2016measuring}. In the chemical system, when taking the response operator $O$ to be the transition dipole operator, $I\left( \omega \right)$ becomes the molecular absorption $A\left(\omega\right)$ which measures the dipole correlation. This shows that, in principle, absorption spectroscopy can measure QFI for a thermal state and thus detect multipartite entanglement. Note that in this approach one must use the full Hamiltonian $H$ which may involve degrees of freedom that the perturbation $O$ does not couple to.

\textit{Squeezed Dicke model.}--- We next consider this generalized QFI inequality for cavity QED, to see how ultra-strong light-matter coupling affects molecular entanglement. Beyond the Tavis-Cummings model  \cite{jaynes1963comparison} that is only valid in the weak coupling regime for polariton chemistry, we generally study the whole coupling regime between a monochromatic cavity mode and $N_{\mathrm{B}}$ molecular excitons (Fig.~\ref{fig:experiments}(a)) described by the Dicke model $H_{\mathrm{D}}$ with the diamagnetic term $H_{\mathrm{A}}$ theoretically scaled by a dimensionless photon squeezing parameter $\kappa\geq 0$, known as the squeezed Dicke model: $H^{\left( \kappa \right)}=H_{\mathrm{D}}+\kappa H_{\mathrm{A}}$, where
\begin{align}
    & H_{\mathrm{D}} = \omega _{\mathrm{c}}a^{\dagger}a+\omega _{\mathrm{m}}\left( S^{\mathrm{z}}+\frac{N_{\mathrm{B}}}{2} \right) +\frac{2G}{\sqrt{N_{\mathrm{B}}}}\left( a^{\dagger}+a \right) S^{\mathrm{x}}
    \label{eq:dicke}
    \\
    & H_{\mathrm{A}} = \frac{G^2}{\omega _{\mathrm{m}}}\left( a^{\dagger}+a \right) ^2~.
    \label{eq:diamagnetic}
\end{align}
Here, $a$ is the field operator (subsystem A) of the cavity mode $\omega_{\mathrm{c}}$. Each molecule is described by a two-level system with singlet states $|\mathrm{g}\rangle$ and $|\mathrm{e}\rangle$ gapped by $\omega_{\mathrm{m}}$, represented by Pauli matrices. The global operators for $N_\mathrm{B}$ molecules (subsystem B) can be written in the Dicke bases as $\vec{S}=\frac{1}{2}\sum_{i=1}^{N_{\mathrm{B}}}{\vec{\sigma}_i}$. $H^{\left( \kappa \right)}$ has the $\mathbb{Z} _2$ parity symmetry $\mathcal{P} =e^{i\pi \left( a^{\dagger}a+S^{\mathrm{z}} \right)}$.  Note that we assume that the cavity-molecule coupling is uniform over all molecules.

In the $N_B\rightarrow\infty$ thermodynamic limit, $H^{\left( \kappa \right)}$ has the feature of superradiant phase transition (SRPT) \cite{emary2003quantum, emary2003chaos, lambert2005entanglement, lambert2004entanglement}. We take a classical mean-field ground state ansatz $|\mathrm{GS}_{\mathrm{cl}}\rangle =|\alpha \rangle _{\mathrm{A}}\otimes \left( \left( -\sin \frac{\theta}{2}|\mathrm{e}\rangle +\cos \frac{\theta}{2}|\mathrm{g}\rangle \right) ^{\otimes N_{\mathrm{B}}} \right) _{\mathrm{B}}$, a product of an oscillator coherent state and a spin coherent state. The classical potential $E_{\mathrm{cl}}^{\left( \kappa \right)}\left( \alpha ,\theta\right) =\langle \mathrm{GS}_{\mathrm{cl}}|H^{\left( \kappa \right)}|\mathrm{GS}_{\mathrm{cl}}\rangle $ has the minima at
\begin{align}
\cos \theta 
&=
\begin{cases}
	1,&\qquad G\leqslant G_{\mathrm{c}}\\
	\kappa +\frac{\omega _{\mathrm{c}}\omega _{\mathrm{m}}}{4G^2}, &\qquad G>G_{\mathrm{c}}\\
\end{cases}
\label{eq:costheta}
\\
\alpha 
&=
\frac{\omega _{\mathrm{m}}}{4G}\sqrt{N_{\mathrm{B}}}\tan \theta~.
\label{eq:alpha}
\end{align}
In the normal phase (disordered phase, $G<G_\mathrm{c}$), the vacuum state with no photon occupation $\left(\alpha=0\right)$ and no molecular excitation $\left(\theta=0\right)$ minimizes the energy. When $G>G_\mathrm{c}$, the system is in the superradiant phase (ordered phase), where the ground state contains a macroscopic number of photons and molecular excitations. Spontaneous $\mathbb{Z} _2$ symmetry breaking occurs with a pair of opposite potential minima, $\left( \theta _{\mathrm{L}},\alpha  _{\mathrm{L}} \right) $ and $\left( \theta  _{\mathrm{R}},\alpha  _{\mathrm{R}} \right)  = \left( -\theta  _{\mathrm{L}},-\alpha  _{\mathrm{L}} \right) $.

For $\kappa <1$ the critical coupling strength is
\begin{equation}
    G_{\mathrm{c}}=\sqrt{\frac{\omega _{\mathrm{m}}\omega _{\mathrm{c}}}{4\left( 1-\kappa \right)}}~.
\end{equation}
Because $\langle H_{\mathrm{A}}\rangle\sim\alpha^2$ has its minimum at $\alpha=0$, increasing $\kappa$ increases $G_{\mathrm{c}}$ and suppresses SRPT. $G_{\mathrm{c}}$ is not well-defined and SRPT cannot occur if $\kappa \geqslant \kappa_{\mathrm{c}} = 1$ (known as SRPT no-go theorem \cite{bialynicki1979no}), which explains the absence of superradiance in molecular polariton experiments. In particular, $\kappa=1$ refers to minimal coupling in the experiments where charges of the particles interact with the cavity modes in the dipole approximation, while $\kappa<1$, including the Dicke model $H^{\left( 0 \right)}$, may be theoretically considered as an effective model to describe SRPT.

To include quantum fluctuation at the classical potential minimum, $H^{\left( \kappa \right)}$ can be expanded in the shifted bases by Holstein–Primakoff transformation \cite{holstein1940field} and Bogoliubov transformation \cite{derezinski2017bosonic}, resulting in of two independent bosonic modes of quasi-particles, known as the upper and lower polaritons: $H^{\left( \kappa \right)}=\Omega _{+}^{\prime}{c_{+}^{\prime}}^{\dagger}c_{+}^{\prime}+\Omega _{-}^{\prime}{c_{-}^{\prime}}^{\dagger}c_{-}^{\prime}+E_{0_+,0_-}$. The polariton modes and zero-point energy are 
\begin{align}
& \Omega _{\pm}^{\prime}=\sqrt{\frac{\tilde{\omega}_{\mathrm{c}}\omega _{\mathrm{c}}+\tilde{\omega}_{\mathrm{m}}^{2}}{2}\pm \sqrt{\left( \frac{\tilde{\omega}_{\mathrm{c}}\omega _{\mathrm{c}}-\tilde{\omega}_{\mathrm{m}}^{2}}{2} \right) ^2+\frac{4G^2\omega _{\mathrm{c}}\omega _{\mathrm{m}}^{2}}{\tilde{\omega}_{\mathrm{m}}}}}
\label{eq:PolaritonEnergy}
\\
& E_{0_+,0_-}=\frac{\Omega _{+}^{\prime}+\Omega _{-}^{\prime}-\omega _{\mathrm{c}}-\tilde{\omega}_{\mathrm{m}}}{2}-N_{\mathrm{B}}\tilde{\omega}_{\mathrm{m}}\sin ^4\frac{\theta}{2}~,
\label{eq:ZeroPointEnergy}
\end{align}
where $\tilde{\omega}_{\mathrm{c}}=\omega _{\mathrm{c}}+4\kappa \frac{G^2}{\omega _{\mathrm{m}}}$ and $\tilde{\omega}_{\mathrm{m}}=\frac{\omega _{\mathrm{m}}}{\cos \theta}$. Under the rotating wave approximation $\left(\left| \omega _{\mathrm{c}}-\omega _{\mathrm{m}} \right| \sim G \ll \omega _{\mathrm{c}}+\omega _{\mathrm{m}}\right)$, Eq.~(\ref{eq:PolaritonEnergy}) becomes $\Omega _{\pm}^{\prime}\approx \omega _{\mathrm{m}}\pm G$, recovering the Tavis-Cummings model. We are interested in the eigenstates $|n_+,n_-\rangle$ and the thermal states $\rho \left( T \right) \propto e^{-H^{\left( \kappa \right)}/T}$. Note that room temperature is much smaller than the electronic energy, so the impact of thermal fluctuation on ground state SRPT is negligible, i.e. $G_{\mathrm{c}}\left( T \right) \approx G_{\mathrm{c}}$.

In the superradiant phase for $N_B\rightarrow\infty$, due to spontaneous symmetry breaking \cite{emary2003chaos},  each eigenenergy becomes two-fold degenerate, and the total Hilbert space is the direct sum of the two asymmetric subspaces $\left(\mathcal{H} =\mathcal{H} _{\mathrm{L}}\oplus \mathcal{H} _{\mathrm{R}}\right)$. The state of interest is the equally-weighted direct sum $\rho  =\frac{1}{2}\rho _{\mathrm{L}} \oplus \frac{1}{2}\rho _{\mathrm{R}} $, which represents an experiment where the two symmetry-broken states each occur with probability one-half.

\begin{figure}[htbp]
    \centering
    \includegraphics[width=0.48\textwidth]{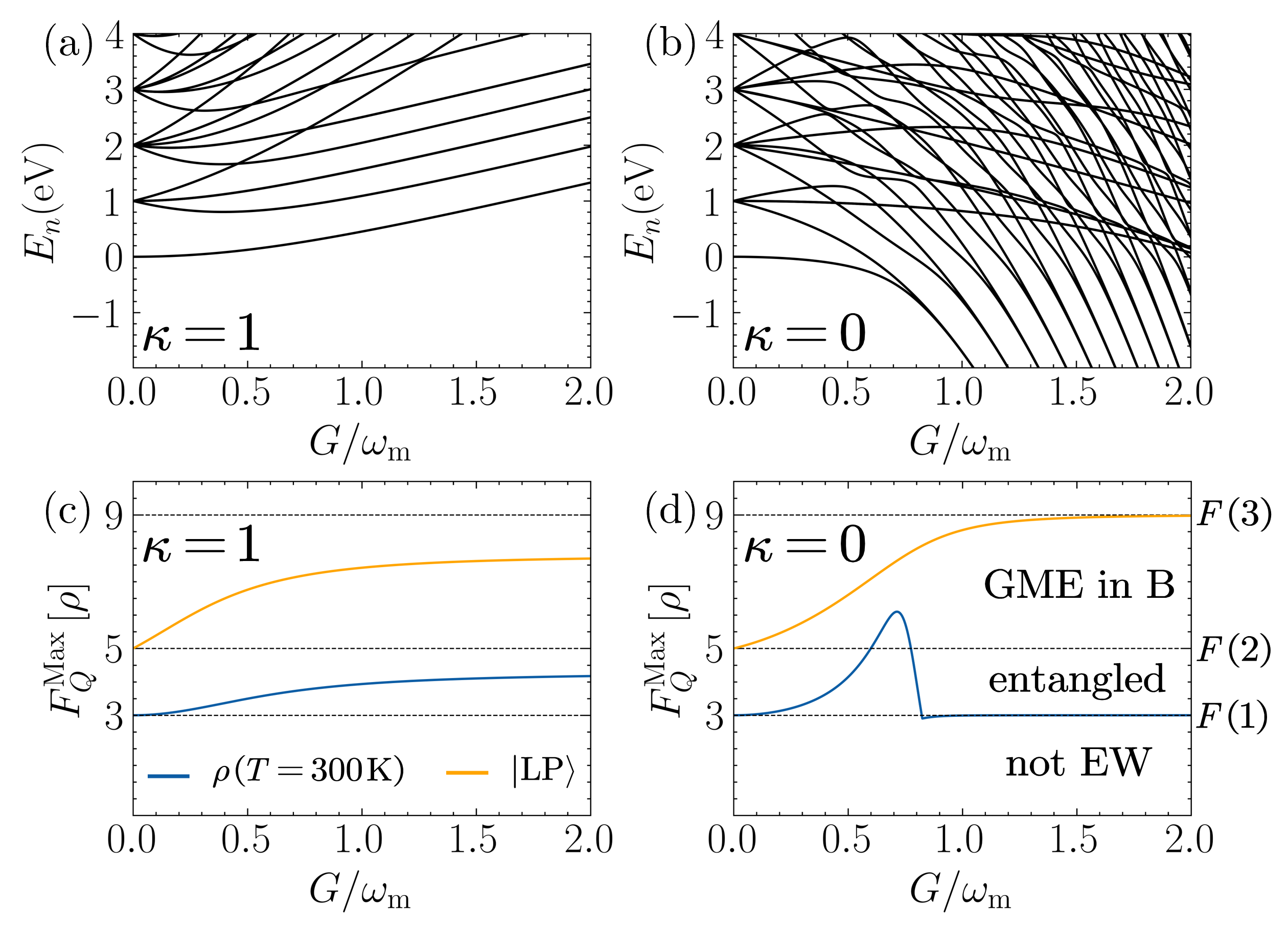}
    \caption{(a)(b) Eigenspectrum $E_n$ as a function of $G$ for minimal coupling model $H^{\left( 1 \right)}$ and Dicke model $H^{\left( 0 \right)}$, for $N_B=3$. (c)(d) Maximized QFI for the room-temperature thermal state and the single lower polariton (the first excited eigenstate) as a function of $G$ for $H^{\left( 1 \right)}$ and $H^{\left( 0 \right)}$. The dashed horizontal lines are the thresholds $F\left(K\right)$ for EW. Parameters: $\omega_{\mathrm{m}}=\omega_{\mathrm{c}}=1~\si{eV}$. Photon number truncation is 70.}
    \label{fig:ThreeMolecules}
\end{figure}

\textit{QFI inequality for Squeezed Dicke model.}---The squeezed Dicke model has $N_{\mathrm{B}}$ molecules plus one cavity mode. To study the multi-molecular entanglement in the squeezed Dicke model, we take $O=\vec{n}\cdot \vec{S}$, where $\vec{n}$ is a unit vector. The spectrum width is $\Delta _i  =1 $ for all molecules. The QFI upper bound function Eq.~(\ref{eq:QFIupperbound}) then is $F\left( K \right) =	q_{\mathrm{B}}K^2+r_{\mathrm{B}}^2$. 
Based on Eq.~(\ref{eq:EWfunctional}), $F_{\mathrm{Q}}\left[ \rho ,\vec{n}\cdot \vec{S} \right] >F\left( k \right) $ means molecular entanglement depth $K\geqslant k+1$. 
We can further optimize $O$ to maximize the QFI to get the best EW, i.e. $F_{\mathrm{Q}}^{\mathrm{Max}}\left[ \rho \right] =\underset{\vec{n}}{\max}\left\{ F_{\mathrm{Q}}\left[ \rho ,\vec{n}\cdot \vec{S} \right] \right\} $. 

We use exact diagonalization to calculate the QFI of three molecules coupled to a cavity, which is the simplest case of multipartite entanglement. 
For the minimal coupling model $H^{\left( 1 \right)}$, 
Fig.~\ref{fig:ThreeMolecules}(c) shows that the room-temperature thermal state is entangled ($F_{\mathrm{Q}}>F\left(1\right)=3$) and the single lower polariton (first excited eigenstate) is genuine multipartite entangled ($F_{\mathrm{Q}}>F\left(2\right)=5$). For both states, the QFI increases with increasing $G$.

The eigenspectrum of the Dicke model $H^{\left( 0 \right)}$  (Fig.~\ref{fig:ThreeMolecules}(b)) shows two-fold near-degeneracy at large $G$, reflecting the superradiant phase.
Fig.~\ref{fig:ThreeMolecules}(d) shows that QFI is a GME witness for the lower polariton and asymptotically reaches the upper bound $F\left(3\right)=9$. For the room-temperature thermal state, $F_{\mathrm{Q}}^{\mathrm{Max}}$ is not monotonic 
but shows a peak because near-degeneracy makes the thermal state becomes more mixed due to the near-degeneracy, which decreases the QFI at large $G$. So QFI can witness thermal state molecular entanglement in the weak or ultrastrong coupling regime at room temperature, and even GME in a certain range of $G$ around the peak, but fails at the deep-strong coupling regime. There is a sharp change in the uninteresting regime ($F_{\mathrm{Q}}^{\mathrm{Max}}<F\left(1\right)=3$) because of the change of the optimized response operator from $S^{\mathrm{x}}$ at smaller $G$ to $S^{\mathrm{y}}$ at larger $G$.

\begin{figure*}[htbp]
    \centering
    \includegraphics[width=1\textwidth]{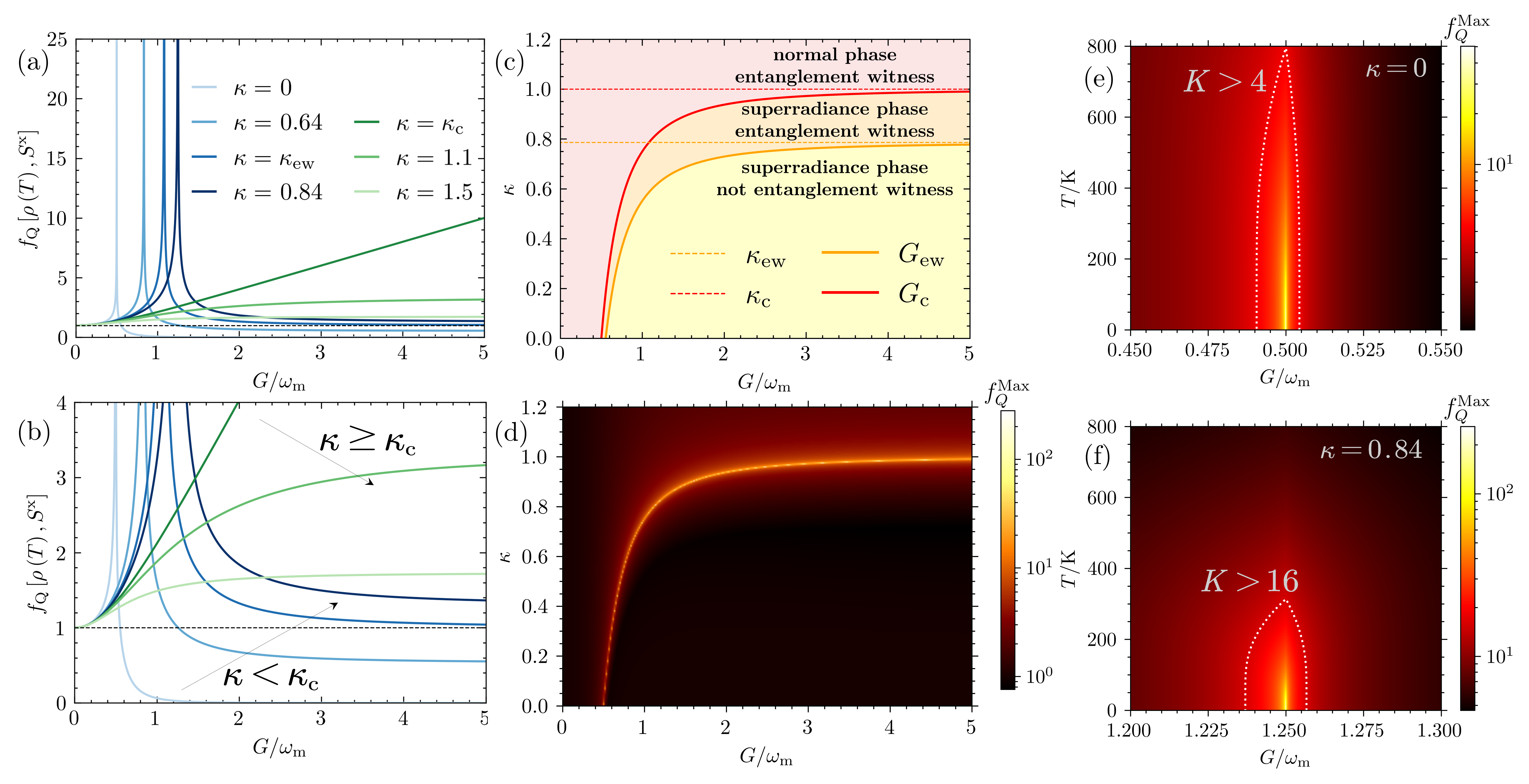}
    \caption{QFI for thermal states of infinite system. (a) $f_{\mathrm{Q}}\left[ \rho \left( T=0\mathrm{K} \right) ,S^{\mathrm{x}} \right]$ as a function of $G$ for different $\kappa$ and (b) the zoomed-in figure with the arrows representing the trends with increasing $\kappa$. (c) Zero-temperature phase diagram. Solid lines are $G_{\mathrm{c}}$ and $G_{\mathrm{ew}}$ as a function of $\kappa$ that divides the figure into three areas. Dashed lines are the asymptotes at $\kappa_{\mathrm{c}}=1$ and $\kappa _{\mathrm{ew}}=\sqrt{\left( \sqrt{5}-1 \right) /2}$, above which $G_{\mathrm{c}}$ and $G_{\mathrm{ew}}$ are not well defined. (d) $f_{\mathrm{Q}}^{\mathrm{Max}}\left[ \rho \left( T=0\mathrm{K} \right)\right]$ as a function of $G$ and $\kappa$ 
    (cutoff around 110). (e) and (f) show nonzero temperatures: $f_{\mathrm{Q}}^{\mathrm{Max}}\left[ \rho \left( T \right) \right]$ as a function of $G$ and $T$ in the density profile with certain cutoff. The area enclosed by white dotted curves represents the validity of multi-molecular entanglement witness for certain entanglement depth $K$. (e) $\kappa=0$, $G_{\mathrm{c}}/\omega_{\mathrm{m}}=1/2$ and $K>4$, (f) $\kappa=0.84$, $G_{\mathrm{c}}/\omega_{\mathrm{m}}=5/4$ and $K>16$. Parameters: $\omega_{\mathrm{m}}=\omega_{\mathrm{c}}=1~\si{eV}$.}
    \label{fig:QFIThermal}
\end{figure*}

\textit{QFI inequality in the thermodynamic limit.}---In the thermodynamic limit, it is convenient to define QFI per molecule $f_{\mathrm{Q}}\left[ \rho ,O \right] ={F_{\mathrm{Q}}\left[ \rho ,O \right]}/{N_{\mathrm{B}}}$, while the EW criterion is $K\geqslant \lfloor f_{\mathrm{Q}}^{\mathrm{Max}} \rfloor +1$. In Supplemental Material \cite{SupplementalMaterial}, we prove that $f_{\mathrm{Q}}^{\mathrm{Max}}\left[ \rho \right] =\underset{\gamma =\mathrm{x},\mathrm{y},\mathrm{z}}{\max}\left\{ f_{\mathrm{Q}}\left[ \rho ,S^{\gamma} \right] \right\}$ for eigenstates and thermal states. 
$f_{\mathrm{Q}}\left[ \rho ,S^{\mathrm{x}} \right]$ is of particular interest, which usually maximizes QFI in the effective EW regime because the light-matter interaction involves $S^{\mathrm{x}}$ explicitly and 
anti-squeezes the system in the x-direction \cite{wang2014quantum}. $S^{\mathrm{x}}$ is proportional to the transition dipole, which relates QFI to the absorption response. By introducing a ``$W$-factor'' as $W_{\pm}=\tanh \left( \frac{\Omega _{\pm}^{\prime}}{2T} \right) $ for $\rho \left( T \right)$ and $W_{\pm}=2n_{\pm}+1$ for $|n_+,n_-\rangle$ with finite $n_{\pm}$,
QFI can be analytically expressed as 
\begin{equation}
f_{\mathrm{Q}}\left[ \rho ,S^{\mathrm{x}} \right] =\omega _{\mathrm{m}}\left( \frac{W_+}{\Omega _{+}^{\prime}}\sin ^2\frac{\zeta}{2}+\frac{W_-}{\Omega _{-}^{\prime}}\cos ^2\frac{\zeta}{2} \right) \cos \theta ~,
\label{eq:fQSx}
\end{equation}
where $\zeta =\arccos \frac{\tilde{\omega}_{\mathrm{c}}\omega _{\mathrm{c}}-\tilde{\omega}_{\mathrm{m}}^{2}}{\sqrt{16G^2\omega _{\mathrm{c}}\omega _{\mathrm{m}}^{2}/\tilde{\omega}_{\mathrm{m}}+\left( \tilde{\omega}_{\mathrm{c}}\omega _{\mathrm{c}}-\tilde{\omega}_{\mathrm{m}}^{2} \right) ^2}}$. 

The zero-temperature QFI in the thermodynamic limit is shown in Fig.~\ref{fig:QFIThermal}(a). If $\kappa<\kappa_{\mathrm{c}}$, QFI is $\lambda$-shaped (blue curves) due to Dicke state squeezing. In the normal phase, with $G$ increasing, the Dicke state is anti-squeezed in the x-direction, which increases $f_{\mathrm{Q}}\left[ \rho ,S^{\mathrm{x}} \right]$. In the superradiant phase, increasing $G$ pulls the two Dicke states far away in the x-direction and less anti-squeezed in each subspace, which decreases $f_{\mathrm{Q}}\left[ \rho ,S^{\mathrm{x}} \right]$. If $\kappa \ge \kappa_{\mathrm{c}}$ and the system is in the normal phase for all $G$,  QFI is monotonic increasing (green curves). QFI may fail to be an EW in the superradiant phase (Fig.~\ref{fig:QFIThermal}(b)). The maximum $G$ where QFI witnesses entanglement is denoted by 
$G_{\mathrm{ew}}$ (Fig.~\ref{fig:QFIThermal}(c) solid orange line), which is not well-defined if $\kappa\ge \kappa_{\mathrm{ew}}= \sqrt{\left( \sqrt{5}-1 \right) /2}\approx0.786$.
In the zero-temperature QFI phase diagram (Fig.~\ref{fig:QFIThermal}(c)), $G_{\mathrm{c}}$ is the critical line for SRPT while $G_{\mathrm{ew}}$ is the boundary for the validation of EW. If $\kappa\ge\kappa_{\mathrm{ew}}$, QFI is an EW for all coupling $G$. If $\kappa<\kappa_{\mathrm{ew}}$, QFI is an EW in the normal phase or near the critical point $G_{\mathrm{c}}$ (up to $G_{\mathrm{ew}}$) in the superradiant phase.  

Zero-temperature QFI (Fig.~\ref{fig:QFIThermal}(c)(d)) dominates when 
$\Omega _{\pm}^{\prime} \gg T$. However, nonzero-temperature becomes substantial near the critical point. The critical behaviours of QFI for zero and nonzero temperature 
are (for $\kappa<1$):
\begin{gather}
\left. f_{\mathrm{Q}}^{\mathrm{Max}}\left[ \rho \left( T=0\right) \right] \right|_{G\rightarrow G_{\mathrm{c}}}\propto \left| G-G_{\mathrm{c}} \right|^{-1/2} ~,
\label{eq:zeroTQFI}
\\
\left. f_{\mathrm{Q}}^{\mathrm{Max}}\left[ \rho \left( T \right) \right] \right|_{G\rightarrow G_{\mathrm{c}}(T)} \approx 
\frac{\tilde{r}^2}{\tilde{r}^2+1}\frac{\omega _{\mathrm{m}}}{2T} ~,
\label{eq:finiteTQFI}
\end{gather}
where 
$\tilde{r}=\frac{\omega _{\mathrm{c}}/\omega _{\mathrm{m}}}{\sqrt{1-\kappa}}$. At $T=0$, SRPT occurs with the signature that $f_{\mathrm{Q}}^{\mathrm{Max}}$ diverges (in the order of $O\left( N_{\mathrm{B}} \right) $, Fig.~\ref{fig:QFIThermal}(a) blue peaks) under the scaling power law at the critical point $G_{\mathrm{c}}$, because QFI has a strong relation to the response correlation function that shows singularity. 
Thus, zero-temperature QFI can be a molecular GME witness at $G_{\mathrm{c}}$: the peak (bright line in Fig.~\ref{fig:QFIThermal}(d)) 
matches $G_{\mathrm{c}}$ in Fig.~\ref{fig:QFIThermal}(c).  At nonzero temperature 
the $\lambda$-peak is capped by $\frac{\omega _{\mathrm{m}}}{T}$. Fig.~\ref{fig:QFIThermal}(e)(f) shows that the divergence of QFI only exists at zero temperature and breaks down dramatically at finite temperature on the order of $T^{-1}$. The nonzero temperature effect makes QFI a weaker multipartite entanglement witness for limited entanglement depth, as shown in the dotted boundaries with small enclosed areas in Fig.~\ref{fig:QFIThermal}(e)(f).

\begin{figure}[htbp]
    \centering
    \includegraphics[width=0.48\textwidth]{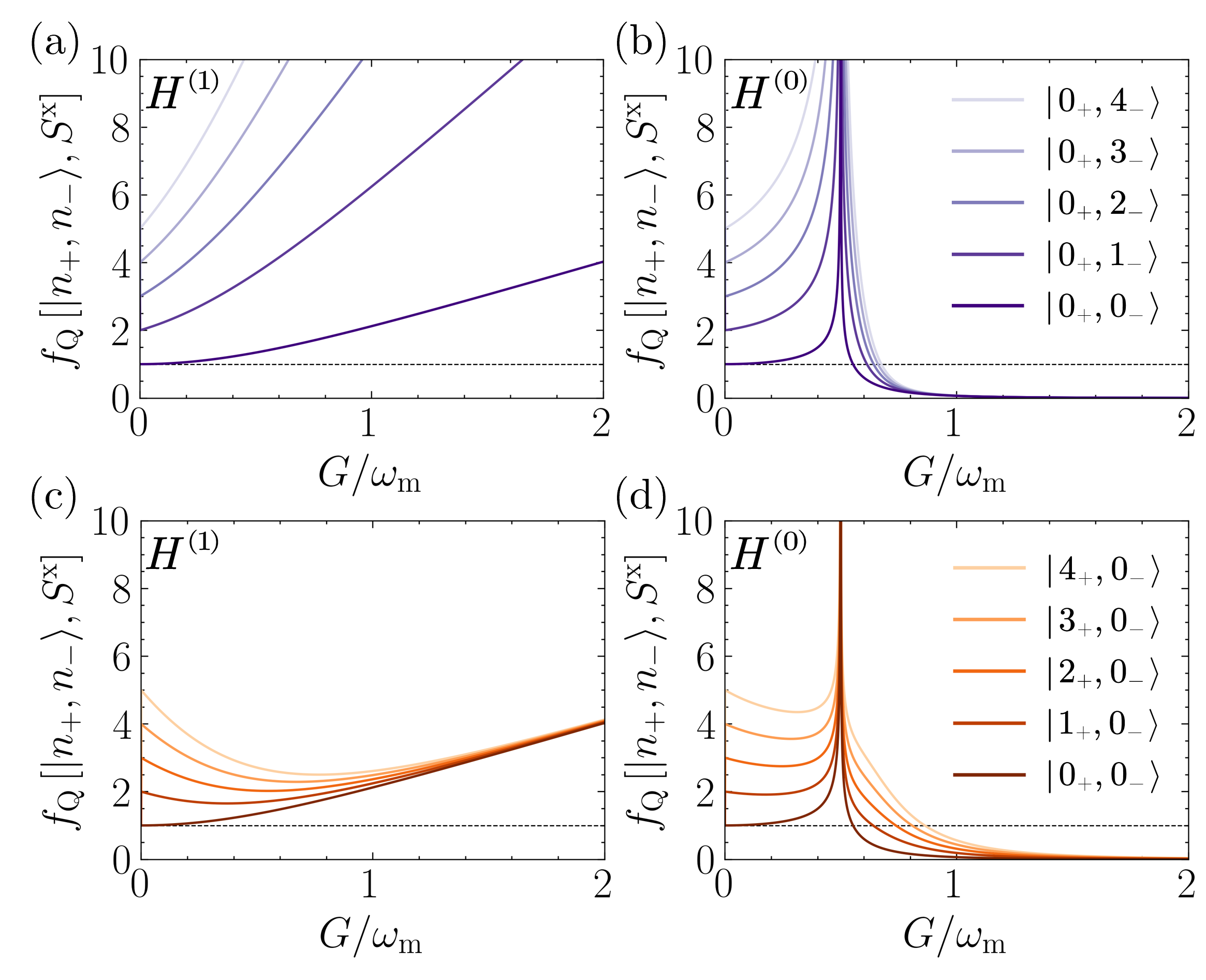}
    \caption{QFI for pure eigenstates for $N_B\rightarrow\infty$, focusing on the upper polariton branch ($\Omega_{+}$, purple) and the lower polariton branch ($\Omega_{-}$, orange). (a), (c) Minimal coupling model without phase transition ($\kappa=1 \ge \kappa_{\mathrm{c}}$). (b), (d) Dicke model with phase transition ($\kappa=0 < \kappa_{\mathrm{c}}$). } 
    \label{fig:QFIeigen}
\end{figure}

QFI for eigenstates can similarly be calculated by Eq.~(\ref{eq:fQSx}), where QFI is linearly dependent on the occupation numbers of both polariton modes $n_+$ and $n_-$. Thus, QFI inequality can better witness larger entanglement depth for higher polariton-occupied states. In Fig.~\ref{fig:QFIeigen}, we focus on the lower polariton branch $|0_+,n_-\rangle $ and the upper polariton branch $|n_+,0_-\rangle $. The trends of QFI for polaritons and QFI for thermal states are similar, showing $\lambda$-shaped singularity when $\kappa<\kappa_{\mathrm{c}}$ and generally increasing when $\kappa\ge\kappa_{\mathrm{c}}$. However, for the upper polariton branch, the QFI decreases at small G before increasing in the normal phase (Fig.~\ref{fig:QFIeigen}(c)(d)), implying a difference between ground state squeezing (coherent spin squeezing) and excited state squeezing.

\textit{QFI measurement for cavity QED.}---Different from the traditional absorption experiments \cite{herrera2017dark} where the polariton absorption is deduced from reflection and transmission, in order to interact with molecules only and probe dipoles directly, bounded mode absorption is required (Fig.~{\ref{fig:experiments}}(b)), where the cavity mode reflection and transmission are attenuated by total internal reflection \cite{herrera2017absorption}. The bounded mode absorption spectrum results from the response of the dipole operator \cite{herrera2017absorption} only
\begin{equation}
A_{\mathrm{b}}\left( \omega \right) \propto \sum_{l,l^{\prime}}{\frac{p_l\left| \langle l|2\mu S^{\mathrm{x}}|l^{\prime}\rangle \right|^2\left( \Gamma _{l^{\prime}}/2 \right)}{\left( \omega +\omega _l-\omega _{l^{\prime}} \right) ^2+\left( \Gamma _{l^{\prime}}/2 \right) ^2}}
\label{eq:BoundedAbsorption}
\end{equation}
where $2\mu S^{\mathrm{x}}$ is the transition dipole operator and the $\delta$-function in Eq.~(\ref{eq:Spectrumfromcorrelation}) is replaced by a Lorentzian of the polariton decay rate $\Gamma _{l^{\prime}}$. QFI for bare molecules is $F_{\mathrm{Q}}\left[ \rho \left( T \right) ,S^{\mathrm{x}} \right] \mid _{G\rightarrow 0}^{}=N_{\mathrm{B}}$, which implies that the QFI per molecule can be calculated by normalizing Eq.~(\ref{eq:QFIfromSpectrum}) 
\begin{equation}
f_{\mathrm{Q}}\left[ \rho \left( T \right) ,S^{\mathrm{x}} \right] =\frac{\int_0^{\infty}{d\omega \frac{\left( 1-e^{-\omega /T} \right) ^2}{1+e^{-\omega /T}}A_{\mathrm{b}}\left( \omega ;T \right)}}{\int_0^{\infty}{d\omega \frac{\left( 1-e^{-\omega /T} \right) ^2}{1+e^{-\omega /T}}A_{\mathrm{b}}\left( \omega ;T \right) \mid _{G\rightarrow 0}^{}}}~.
\label{eq:fQfromabsorption}
\end{equation}
The magnitude of transition dipole $\mu$ and other unknown coefficients are all removed by normalization. For the polariton states $|n_+,n_-\rangle$, the measurement of variance can be deduced from the pump-probe spectroscopy (excited state absorption)
\begin{equation}
\frac{f_{\mathrm{Q}}\left[ |n_+,n_-\rangle ,S^{\mathrm{x}} \right]}{n_++n_-+1}=\frac{\int_0^{\infty}{d\omega A_{\mathrm{b}}\left( \omega ;n_+,n_- \right)}}{\int_0^{\infty}{d\omega A_{\mathrm{b}}\left( \omega ;n_+,n_- \right) \mid _{G\rightarrow 0}^{}}}~,
\label{eq:fQfromTA}
\end{equation}
where $\langle n_+,n_-|S^{\mathrm{x}}|n_+,n_-\rangle =0$ due to $\mathbb{Z} _2$ symmetry.
The most easily detected polariton states in the pump-probe experiments are $|1_+,0_-\rangle$ and $|0_+,1_-\rangle$.

Eq.~(\ref{eq:fQfromabsorption}) and Eq.~(\ref{eq:fQfromTA}) offer an experimental protocol to detect entanglement in polariton chemistry: measuring the absorption of a bare sample and the bounded mode absorption of another sample inside the cavity. Calculate QFI so that the entanglement depth $K$ of the thermal system satisfies $K\geqslant \lfloor f_{\mathrm{Q}} \rfloor +1$.

\textit{Conclusion and outlook.}---To summarize, we proposed an experimentally feasible protocol based on QFI that can sufficiently detect entanglement in chemical systems. We generalize the QFI inequality from identical \cite{hyllus2012fisher} to non-identical local response operators to fit various particles in one chemical system. We use the generalized QFI inequality as a multipartite entanglement witness, to reveal if chemical systems at room temperature can maintain multi-molecular entanglement. We then connect QFI to the dipole correlator, which can be directly measured from the absorption spectrum. This protocol is generally feasible for the multipartite entanglement of a macroscopic number of molecules with multi-excitations 
, such as molecular aggregates \cite{hestand2017molecular}, polaritons \cite{mandal2022theoretical}
and biological proteins 
\cite{lee2007coherence}, to detect if direct or indirect inter-molecular interaction may help to build and maintain long-lived molecular entanglement. If our protocol proves molecular entanglement is long-lived in certain chemical systems, the non-trivial quantum effect of entanglement may explain the coherent delocalization in organic structure or be used as quantum resources for information processing. Our protocol is also a benchmark for exact quantum tomography \cite{lvovsky2009continuous}.

We further tested the proposed entanglement witness theoretically for electronic polariton systems in ultrastrong coupling regime (squeezed Dicke model), where the counter-rotating and the diamagnetic terms \cite{frisk2019ultrastrong} are considered. The analytical results are summarized in the phase diagram (Fig.~\ref{fig:QFIThermal}(c)), where QFI inequality can witness entanglement for thermal states or polariton states for moderate $G$ or large $\kappa$, and witness larger entanglement depth (even GME) around the superradiant critical point due to the singularity of Dicke state anti-squeezing. The chemical systems are usually in the $\kappa \sim 1$ regime in Fig.~\ref{fig:QFIThermal}(c), because the extremely large light-matter coupling is difficult to achieve and the diamagnetic term suppresses SRPT (the no-go theorem \cite{bialynicki1979no}). Therefore, our protocol successfully proves the existence of long-lived multi-molecular entanglement induced by cavity QDE.

We emphasize that our prediction of entanglement for polariton chemistry is experimentally verifiable via ultrafast spectroscopy techniques for bounded mode absorption \cite{herrera2017absorption}, and is unrestricted by the unknown dipole.

It is interesting to consider the chemical significance of cavity-induced entanglement. Strong non-classical correlations between spatially distant sites may demonstrate the physics of polariton-enhanced exciton transport and charge separation \cite{wu2022polariton} with better robustness and efficiency, offering insights for the next generation of organic photovoltaics. Our framework and protocol (Eq.~\ref{eq:EWfunctional} and Eq.~\ref{eq:fQSx}) are also valid for vibrational polaritons. For example, the change of entanglement depth may be beneficial to finding the reaction's active center in cavity-controlled chemical selectivity, when different sites are dressed with different local perturbations. Investigating the thermal fluctuation for non-zero temperature SRPT at the vibrational energy scale is also interesting.

Returning to the foundation of QFI, polariton chemistry may be a platform for quantum-enhanced sensing and metrology with room-temperature molecular sensors via squeezed Dicke states, because entangled states are more sensitive probes of the environment than classical states. It is important to build extreme quantum resources characterized by QFI singularity, where SRPT occurs with a GME state that reaches the Heisenberg limit. Thus, exploring the experimental approach for SRPT in polariton chemistry, such as the anti-squeezing effect via magnetic polaritons \cite{chen2021experimental,bamba2022magnonic}, or out-of-equilibrium states \cite{zhang2021observation, ferioli2023non, baumann2010dicke} like Raman transition coupling \cite{dimer2007proposed, baden2014realization} that can achieve $\kappa<1$, is urged.

\begin{acknowledgments}
W.W. and G.D.S. are funded by the National Science Foundation under Grant No. 2211326. D.A.H. was supported in part by NSF QLCI grant OMA-2120757.  We acknowledge Andrew E. Sifain, Abraham Nitzan, Bing Gu and Ava N. Hejazi for the valuable discussions.
\end{acknowledgments}


\bibliography{apssamp}

\newpage
\clearpage

\pagebreak
\widetext
\begin{center}
\textbf{\large Supplemental Material: Molecular Entanglement Witness by Absorption Spectroscopy in Cavity QED}
\end{center}
\setcounter{equation}{0}
\setcounter{figure}{0}
\setcounter{table}{0}
\setcounter{page}{1}
\makeatletter
\renewcommand{\theequation}{S\arabic{equation}}
\renewcommand{\thefigure}{S\arabic{figure}}
\renewcommand{\bibnumfmt}[1]{[S#1]}
\renewcommand{\citenumfont}[1]{S#1}

\section{Proof for generalized entanglement witness related to QFI}
In this section, we prove the theorems of general entanglement witness related to quantum Fisher information regarding non-identical local response operators.

\begin{definition}[$k$-producible]
    A pure state is $k$-producible if it can be factorized as $|\Phi \rangle =\bigotimes_l{|\phi \rangle _l}$, where each sub-state $|\phi \rangle _l$ is a $N_l$-particle state ($N=\sum_l{N_l}$) with $N_l\leqslant k$. A mixed state is $k$-producible if it can be convexly linear-combined by $k$-producible pure state $\rho =\sum_{\nu}{p_{\nu}|\Phi _{\nu}\rangle \langle \Phi _{\nu}|}$. 
\end{definition}

\begin{definition}[entanglement depth $K$]
    A state of Entanglement depth $K$ means that the state is $K$-producible but not $\left(K-1\right)$-producible.
\end{definition}

\begin{definition}[$k^{\prime}$-producible in $N_{\mathrm{B}}$ particles]
    Consider a system $\rho$ containing $N$ different particles, $N_{\mathrm{B}}$ of which are special. Assume that a $k$-producible pure state can be factorized as as $|\Phi \rangle =\bigotimes_l{|\phi \rangle _l}$, where each sub-state $|\phi \rangle _l$ is a $N_l$-particle state ($N=\sum_l{N_l}$) with $N_l\leqslant k$. For each $|\phi \rangle _l$, $N_{\mathrm{B},l}$ of the $N_l$ particles belong to the $N_{\mathrm{B}}$ particles, with $N_{\mathrm{B},l} \leqslant N_l$ and $N_{\mathrm{B}} = \sum_{l}{N_{\mathrm{B},l}}$. If $N_{\mathrm{B},l} \leqslant k^{\prime} \leqslant k$, the state $\rho$ is called $k^{\prime}$-producible in $N_{\mathrm{B}}$ particles. A mixed state is $k^{\prime}$-producible if it can be convexly linear-combined by $k^{\prime}$-producible pure state $\rho =\sum_{\nu}{p_{\nu}|\Phi _{\nu}\rangle \langle \Phi _{\nu}|}$. Note that ``$k^{\prime}$-producible in $N_{\mathrm{B}}$ particles'' is defined on the full system rather than the reduced density matrix of the $N_{\mathrm{B}}$ particles, i.e. no partial trace is required, which implies the pattern of multipartite entanglement of the full system.
\end{definition}

\begin{definition}[entanglement depth $K^{\prime}$ in $N_{\mathrm{B}}$ particles]
    A state is of ``Entanglement depth $K^{\prime}$ in $N_{\mathrm{B}}$ particles'' means that it is ``$K^{\prime}$-producible in $N_{\mathrm{B}}$ particles'' but not $\left(K^{\prime}-1\right)$-producible in $N_{\mathrm{B}}$ particles''. Note that ``entanglement depth in $N_{\mathrm{B}}$ particles'' is defined on the full system rather than the reduced density matrix of the $N_{\mathrm{B}}$ particles, i.e. no partial trace is required.
\end{definition}

\begin{lemma}
The sets of k-producible states are convex.
\end{lemma}
\begin{proof}
A pure state is $k$-producible if it can be factorized as $|\Phi \rangle =\bigotimes_l{|\phi \rangle _l}$, where each sub-state $|\phi \rangle _l$ is a $N_l$-particle state ($N=\sum_l{N_l}$) with $N_l\leqslant k$. A mixed state is $k$-producible if it can be convexly linear-combined by $k$-producible pure state $\rho =\sum_{\nu}{p_{\nu}|\Phi _{\nu}\rangle \langle \Phi _{\nu}|}$. The definition of $k$-producibility has already contained the convexity.
\end{proof}

\begin{lemma}
    For a constant operator $O$, QFI is convex regarding the eigenstates of the density matrix.
    \begin{equation}
        F_{\mathrm{Q}}\left[ \rho =\sum_i{p_i|i\rangle \langle i|},O \right] \leqslant \sum_i{p_iF_{\mathrm{Q}}\left[ |i\rangle ,O \right]}
    \end{equation}
\end{lemma}
\begin{proof}
    Based on the definition, QFI can be equivalently written as
    \begin{equation}
        F_{\mathrm{Q}}\left[ \rho ,O \right] =4\sum_i^{\mathrm{dim}\left( \rho \right)}{p_i\left( \langle i|O^2|i\rangle -\langle i|O|i\rangle ^2 \right)}-8\sum_{i\ne j}^{\mathrm{dim}\left( \rho \right)}{\frac{p_ip_j}{p_i+p_j}\left| \langle i|O|j\rangle \right|^2}
    \end{equation}
    where the first term is QFI for a pure state $F_{\mathrm{Q}}\left[ |i\rangle ,O \right] =4\left( \langle i|O^2|i\rangle -\langle i|O|i\rangle ^2 \right) $ while the second term is always non-positive. Dropping the second term proves this lemma.
\end{proof}

\begin{lemma}
    For any pure state factorized as $|\Psi \rangle =\bigotimes_{l=1}^M{|\Psi \rangle _l}$, and the response operator is the sum of local operators $O=\sum_{l=1}^M{O_l}$, where $|\Psi \rangle _l$ and $O_l$ are in the same sub Helbert space $\mathcal{H} _l$, QFI obeys the sum rule
    \begin{equation}
        F_{\mathrm{Q}}\left[ |\Psi \rangle ,O \right] =\sum_{l=1}^M{F_{\mathrm{Q}}\left[ |\Psi \rangle _l,O_l \right]}
    \end{equation}
\end{lemma}
\begin{proof}
    Because QFI of a pure state is proportional to the variance, and each subspace $\mathcal{H} _l$ is independent of each other, this lemma is the same as the sum rule for the variance of independent variables.
\end{proof}

\begin{lemma}
    Given a constant operator $O$ with the spectrum $O|\Lambda _i\rangle =\Lambda _i|\Lambda _i\rangle$, and the spectrum range $\Lambda _{\mathrm{M}}=\max \left\{ \Lambda _i \right\} , \Lambda _{\mathrm{m}}=\min \left\{ \Lambda _i \right\}$, the upper bound of QFI is
    \begin{equation}
        F_{\mathrm{Q}}\left[ \rho ,O \right] \leqslant \left( \Lambda _{\mathrm{M}}-\Lambda _{\mathrm{m}} \right) ^2
    \end{equation}
\end{lemma}

\begin{proof}
    Because $F_{\mathrm{Q}}\left[ \rho =\sum_i{p_i}|\psi _i\rangle \langle \psi _i|,O \right] \leqslant \sum_i{p_iF_{\mathrm{Q}}\left[ |\psi _i\rangle \langle \psi _i|,O \right]}\leqslant \underset{|\psi _i\rangle}{\max}\left\{ F_{\mathrm{Q}}\left[ |\psi _i\rangle \langle \psi _i|,O \right] \right\} $, for any given mixed state QFI, there is always a pure state QFI no less than it. This means the QFI can be maximized by a pure state $\underset{\rho}{\max}\left\{ F_{\mathrm{Q}}\left[ \rho ,O \right] \right\} =\underset{|\psi \rangle}{\max}\left\{ F_{\mathrm{Q}}\left[ |\psi \rangle ,O \right] \right\} $ and the veriance is upper bounded by $F_{\mathrm{Q}}\left[ |\psi \rangle ,O \right] =4\mathrm{Var}\left( |\psi \rangle ,O \right) \leqslant \left( \Lambda _{\mathrm{M}}-\Lambda _{\mathrm{m}} \right) ^2$. The equality condition for the upper bound is $|\psi \rangle =\frac{1}{\sqrt{2}}\left( |\Lambda _{\mathrm{M}}\rangle +e^{i\phi}|\Lambda _{\mathrm{m}}\rangle \right)$ with the equally-weighted coherent superposition of $|\Lambda _{\mathrm{m}}\rangle $ and $|\Lambda _{\mathrm{M}}\rangle $ up to a relative phase.
\end{proof}

\begin{theorem}[Upper Bound of QFI]
    Consider a system $\rho$ containing $N$ different particles, each with their own different local operator $\left\{ O_i \right\} $. The total response operator is $O=\sum_{i=1}^N{O_i}$. Each local operator has its spectrum $O_i|\lambda _{l,i}\rangle =\lambda _{l,i}|\lambda _{l,i}\rangle$. Without loss of generality, relabel the particles in descending order of local operator spectrum width, $\Delta _1\geqslant \Delta _2\geqslant \cdots \geqslant \Delta _N$, where $\Delta _i=\underset{l}{\max}\left\{ \lambda _{l,i} \right\} -\underset{l}{\min}\left\{ \lambda _{l,i} \right\} $. If $N$-particle state $\rho$ is of entanglement depth $K$, QFI is upper bounded by
    \begin{equation}
        F_{\mathrm{Q}}\left[ \rho ,O \right] \leqslant F\left( K;\left\{ \Delta _i \right\} \right) =\sum_{l=1}^q{\left( \sum_{j=1}^K{\Delta _{K\left( l-1 \right) +j}} \right) ^2+\left( \sum_{j=1}^r{\Delta _{Kq+j}} \right) ^2}
        \label{eqS:witness}
    \end{equation}
where $q=\lfloor \frac{N}{K} \rfloor$ and $r=N-Kq$ are the quotient and remainder.
\end{theorem}
\begin{proof}
    In Helbert space $\mathcal{H}$, given a state $\rho$ that is $k$-producible, based on Lemma 1 and Lemma 2, the maximum value reaches the $k$-producible pure state.
    \begin{equation}
        \underset{\rho _{k-\mathrm{prod}}}{\max}\left\{ F_{\mathrm{Q}}\left[ \rho _{k-\mathrm{prod}},O \right] \right\} =\underset{|\psi _{k-\mathrm{prod}}\rangle}{\max}\left\{ F_{\mathrm{Q}}\left[ |\psi _{k-\mathrm{prod}}\rangle ,O \right] \right\} 
    \end{equation}
    So we only need to check the pure states to get the upper bound. Suppose the $k$-producible pure state is factorized as $|\Psi _{k\mathrm{-prod}}\rangle =\bigotimes_{l=1}^M{|\psi \rangle _l}, \, N_l\leqslant k$, where each state $|\psi \rangle _l$ lies in a $N_l$-particle subspace $\mathcal{H} _l$. The particles that make up $\mathcal{H} _l$ have their label in the set $\mathcal{S} _l$. So we have $N=\sum_{l=1}^M{N_l},\,\mathcal{H} =\bigotimes_{l=1}^M{\mathcal{H} _l},\,\left\{ 1,...,N \right\} =\bigcup_{l=1}^M{\mathcal{S} _l}$ Thus we can define $O_l$ as the local response operator for each set of particles as $O_l=\sum_{i\in S_l}{O_i}$, where $O=\sum_{l=1}^M{O_l}$. Based on Lemma 3 and Lemma 4, we have 
    \begin{equation}
        F_{\mathrm{Q}}\left[ |\Psi _{k\mathrm{-prod}}\rangle ,O \right] =\sum_{l=1}^M{F_{\mathrm{Q}}\left[ |\psi \rangle _l,O_l=\sum_{i\in \mathcal{S} _l}{O_i} \right]}\leqslant \sum_{l=1}^M{\left( \sum_{i\in \mathcal{S} _l}{\Delta _i} \right) ^2}
    \end{equation}
    
    So we need to maximize $\sum_{l=1}^M{\left( \sum_{i\in \mathcal{S} _l}{\Delta _i} \right) ^2}$ by optimizing $\mathcal{S} _l$, i.e., the way of grouping particles into sets with each set having no more than $k$ particles. The optimal grouping is to put the first $k$ particle in a group $\mathcal{S} _1=\left\{ 1,...,k \right\} $, the second $k$ particle in a group $\mathcal{S} _2=\left\{ k+1,...,2k \right\} $, ..., the rest $r$ particle in a group $\mathcal{S} _{q+1}=\left\{ qk+1,...,qk+r \right\} $. Notice that $\Delta$ has already in the descending order $\Delta _1\geqslant \Delta _2\geqslant \cdots \geqslant \Delta _N$. To prove this, let's first prove the optimal grouping satisfying $i<j$ for any $i \in \mathcal{S} _l, j \in \mathcal{S} _{l+1}$. If this is not the case ($i>j$), we can swap particle $i$ and particle $j$. Due to rearrangement inequality ($AX+BY \ge AY+BX$ if $A\ge B \ge0$ and $X\ge Y \ge0$), the action of swapping increases the value of $\sum_{l=1}^M{\left( \sum_{i\in S_l}{\Delta _i} \right) ^2}$. Next, we prove the number of particles in $\mathcal{S} _1$ is $k$. If this is not the case, we can move the one particle from $\mathcal{S} _2$ to $\mathcal{S} _1$, which also increases the value of $\sum_{l=1}^M{\left( \sum_{i\in S_l}{\Delta _i} \right) ^2}$ because of rearrangement inequality (where $Y=0$). Now we have $\mathcal{S} _1=\left\{ 1,...,k \right\} $. For the rest $N-k$ particles, we can recursively prove $\mathcal{S} _2=\left\{ k+1,...,2k \right\} $, ..., $\mathcal{S} _{q+1}=\left\{ qk+1,...,qk+r \right\} $. So we have 
    \begin{equation}
        F_{\mathrm{Q}}\left[ |\Psi _{k\mathrm{-prod}}\rangle ,O \right] \leqslant \left( \Delta _1+...+\Delta _k \right) ^2+\left( \Delta _{k+1}+...+\Delta _{2k} \right) ^2+...+\left( \Delta _{N-r+1}+...+\Delta _N \right) ^2
        \label{eqS:ExplictUpperBound}
    \end{equation}
    The upper bound is exact $F\left( k;\left\{ \Delta _i \right\} \right) $ with the equality condition 
    \begin{equation}
        |\Psi _{k\mathrm{-prod}}\rangle =\left( \bigotimes_{l=1}^q{|\psi \rangle _l} \right) \otimes |\psi \rangle _{q+1}
    \end{equation}
    where $|\psi \rangle _l$ and $|\psi \rangle _{q+1}$ are the GHZ states
    \begin{gather}
        |\psi \rangle _l=\frac{1}{\sqrt{2}}\left( \bigotimes_{i=1}^k{|\lambda _{\mathrm{M},\left( l-1 \right) k+i}\rangle}+\bigotimes_{i=1}^k{|\lambda _{\mathrm{m},\left( l-1 \right) k+i}\rangle} \right) 
        \\
        |\psi \rangle _{q+1}=\frac{1}{\sqrt{2}}\left( \bigotimes_{i=1}^r{|\lambda _{\mathrm{M},qk+i}\rangle}+\bigotimes_{i=1}^r{|\lambda _{\mathrm{m},qk+i}\rangle} \right) 
    \end{gather}
    We can also see that $F\left( k;\left\{ \Delta _i \right\} \right) $ is a monotonic increasing function regarding $k$ because the grouping method with the constraint $N_l \leqslant k-1$ also a non-optimal grouping method for with the constraint $N_l \leqslant k$. 
    
    We can also change the $k$-producibility into entanglement depth $K$ because producibility is not unique. If a state is $k$-producible, it must be $\left(k+1\right)$-producible. On the contrary, entanglement depth $K$ is defined as the state is $K$-producible but not $\left(K-1\right)$-producible, which is unique. Due to monotonicity, entanglement depth $K$ generates the supremum of $F_{\mathrm{Q}}\left[ \rho ,O \right]$. By replacing $k$ by $K$ in Eq.~(\ref{eqS:ExplictUpperBound}), we prove the theorem Eq.~(\ref{eqS:witness}).
\end{proof}

\begin{theorem}[Upper Bound of QFI for partial perturbation]
    Consider a system $\rho$ containing $N$ different particles, $N_{\mathrm{B}}$ of which are each perturbed by a local operator $\left\{ O_i \right\} $. The total response operator is $O_{\mathrm{B}}=\sum_{i=1}^{N_{\mathrm{B}}}{O_i}$. Each local operator has its spectrum $O_i|\lambda _{l,i}\rangle =\lambda _{l,i}|\lambda _{l,i}\rangle$. Without loss of generality, relabel the particles in descending order of local operator spectrum width, $\Delta _1\geqslant \Delta _2\geqslant \cdots \geqslant \Delta _{N_{\mathrm{B}}}>0$, where $\Delta _i=\underset{l}{\max}\left\{ \lambda _{l,i} \right\} -\underset{l}{\min}\left\{ \lambda _{l,i} \right\} $. If $N$-particle state $\rho$ is of entanglement depth $K\leqslant N_{\mathrm{B}}$, the QFI is upper bounded by 
    \begin{equation}
        F_{\mathrm{Q}}\left[ \rho ,O_{\mathrm{B}} \right] \leqslant 
        F_{\mathrm{B}}\left( K;\left\{ \Delta _i \right\} \right)= 
        \sum_{l=1}^{q_{\mathrm{B}}}{\left( \sum_{j=1}^K{\Delta _{K\left( l-1 \right) +j}} \right) ^2+\left( \sum_{j=1}^{r_{\mathrm{B}}}{\Delta _{Kq_{\mathrm{B}}+j}} \right) ^2}
    \label{eqS:QFIupperboundPartialPerturbation}
    \end{equation}
    where, $q_{\mathrm{B}}=\lfloor \frac{N_{\mathrm{B}}}{K} \rfloor$ and $r_{\mathrm{B}}=N_B-Kq_{\mathrm{B}}$ are the quotient and remainder. 
\end{theorem}

\begin{proof}
    Theorem 2 is a special case of Theorem 1, by taking $\Delta _i=0$ for $i>N_{\mathrm{B}}$. So we have
    \begin{equation}
        \begin{aligned}
            F_{\mathrm{Q}}\left[ \rho ,O \right] 
            \leqslant & 
            \left( \Delta _1+...+\Delta _k \right) ^2+\left( \Delta _{k+1}+...+\Delta _{2k} \right) ^2+...+\left( \Delta _{N-r+1}+...+\Delta _N \right) ^2
            \\
            =&
            \left( \Delta _1+...+\Delta _k \right) ^2..+\left( \Delta _{q_{\mathrm{B}}k+1}+...+\Delta _{N_{\mathrm{B}}}+...+\Delta _{\left( q_{\mathrm{B}}+1 \right) k} \right) ^2+...+\left( \Delta _{N-r+1}+...+\Delta _N \right) ^2
            \\
            =&
            \left( \Delta _1+...+\Delta _k \right) ^2+...+\left( \Delta _{q_{\mathrm{B}}k+1}+...+\Delta _{N_{\mathrm{B}}} \right) ^2
        \end{aligned}
        \label{eqS:ExplictUpperBoundPartialPerturbation}
    \end{equation}
    Thus we may simply replace $q$ and $r$ by $q_{\mathrm{B}}$ and $r_{\mathrm{B}}$ respectively in Eq.~(\ref{eqS:witness}), resulting in Eq.~(\ref{eqS:QFIupperboundPartialPerturbation}). Note that Theorem 1 can be recovered from Theorem 2 by taking $N_{\mathrm{B}} = N$.
\end{proof}

\begin{theorem}[Upper Bound of QFI for partial perturbation with the pattern of multipartite entanglement]
    Consider a system $\rho$ containing $N$ different particles, $N_{\mathrm{B}}$ of which are each perturbed by a local operator $\left\{ O_i \right\} $. The total response operator is $O_{\mathrm{B}}=\sum_{i=1}^{N_{\mathrm{B}}}{O_i}$. If the $N$-particle state $\rho$ is of entanglement depth $K^{\prime}$ in $N_{\mathrm{B}}$ particles ($K^{\prime}\leqslant N_{\mathrm{B}}\leqslant N$), the QFI is upper-bounded by
    \begin{equation}
        F_{\mathrm{Q}}\left[ \rho ,O_{\mathrm{B}} \right] \leqslant 
        F_{\mathrm{B}}\left( K^{\prime};\left\{ \Delta _i \right\} \right)
        \label{eqS:PartialPerturbation}
    \end{equation}
    Here, $F_{\mathrm{B}}\left( k^{\prime};\left\{ \Delta _i \right\} \right)$ is defined in Eq.~(\ref{eqS:QFIupperboundPartialPerturbation}).
\end{theorem}
\begin{proof}
    Assume that a $k$-producible pure state can be factorized as as $|\Phi \rangle =\bigotimes_l{|\phi \rangle _l}$, where each sub-state $|\phi \rangle _l$ is a $N_l$-particle state ($N=\sum_l{N_l}$) with $N_l\leqslant k$. For each $|\phi \rangle _l$, $N_{\mathrm{B},l}$ of the $N_l$ particles belong to the $N_{\mathrm{B}}$ particles, with $N_{\mathrm{B},l} \leqslant N_l$ and $N_{\mathrm{B}} = \sum_{l}{N_{\mathrm{B},l}}$. If there is an additional constraint that for each $|\phi \rangle _l$, the number of particles from $N_{\mathrm{B}}$ particles should be no more than $k^{\prime}$, i.e. $N_{\mathrm{B},l} \leqslant k^{\prime} \leqslant k$, the upper bound of QFI becomes 
    \begin{equation}
        \begin{aligned}
            F_{\mathrm{Q}}\left[ |\Psi _{k-\mathrm{prod}}\rangle ,O_{\mathrm{B}} \right] 
            \leqslant & 
            \sum_{l=1}^M{\left( \sum_{i\in S_l}{\Delta _i} \right) ^2} 
            \\
            \leqslant & 
            \left( \Delta _1+...+\Delta _{k^{\prime}}+0+...+0 \right) ^2+\left( \Delta _{k^{\prime}+1}+...+\Delta _{2k^{\prime}}+0+...+0 \right) ^2+...
        \end{aligned}
    \end{equation}
    where the maximum reaches when in each group $\mathcal{S} _l$ there are $k^{\prime}$ (or less) particles from subsystem B and the rest filled by $i>N_{\mathrm{B}}$. The upper bound can be equivalently written as $F_{\mathrm{B}}\left( k^{\prime};\left\{ \Delta _i \right\} \right)$ by dropping all the 0, which is defined in Eq.~(\ref{eqS:QFIupperboundPartialPerturbation}). Similarly, for the $k$-producible mixed state, we may use Lemma 2 to prove that its QFI have the same upper bound as the $k$-producible pure state
    \begin{equation}
        F_{\mathrm{Q}}\left[ \rho_{k-\mathrm{prod}} ,O_{\mathrm{B}} \right] \leqslant F_{\mathrm{B}}\left( k^{\prime};\left\{ \Delta _i \right\} \right)
    \end{equation}
    Note that in the RHS, there is no $k$-dependence, which implies that the upper bound of QFI is only dependent on the $k^{\prime}$-producibility in $N_{\mathrm{B}}$ particles, even though $\rho_{k-\mathrm{prod}}$ is a state in the full Hilbert space. So we do not need to point out the $k$-producibility in Theorem 3. Similarly, we may replace the ``$k^{\prime}$-producibility in $N_{\mathrm{B}}$ particles" by the unique ``entanglement depth $K^{\prime}$ in $N_{\mathrm{B}}$ particles''. So we prove Eq.~(\ref{eqS:PartialPerturbation}).
\end{proof}

\section{Analytical solution of squeezed Dicke model}

In this section, we analytically solve the squeezed Dicke model with the only assumption of thermodynamics limit ($N_{\mathrm{B}}\rightarrow \infty$), which is easy to accomplish in the macroscopic scale. The squeezed Dicke model is written as
\begin{equation}
  H^{\left(\kappa\right)} = \omega _{\mathrm{c}}a^{\dagger}a+\omega _{\mathrm{m}}\left( S^{\mathrm{z}}+\frac{N_{\mathrm{B}}}{2} \right) +\frac{2G}{\sqrt{N_{\mathrm{B}}}}\left( a^{\dagger}+a \right) S^{\mathrm{x}} + \kappa \frac{G^2}{\omega _{\mathrm{m}}}\left( a^{\dagger}+a \right) ^2
\end{equation}

We take a classical ground state ansatz $|\mathrm{GS}_{\mathrm{cl}}\rangle =|\alpha \rangle _{\mathrm{A}}\otimes \left( \left( -\sin \frac{\theta}{2}|\mathrm{e}\rangle +\cos \frac{\theta}{2}|\mathrm{g}\rangle \right) ^{\otimes N_{\mathrm{B}}} \right) _{\mathrm{B}}$, consisting of an oscillator coherent state $\left(\alpha \in \mathbb{R}\right)$ and a spin coherent state $\left(\theta \in \left[ -\frac{\pi}{2},\frac{\pi}{2} \right] \right)$. The classical potential regarding the coordinates $\theta$ and $\alpha$ is 
\begin{equation}
H_{\mathrm{cl}}^{\left( \kappa \right)}\left( \alpha ,\theta \right) =\langle \mathrm{GS}_{\mathrm{cl}}|H^{\left( \kappa \right)}|\mathrm{GS}_{\mathrm{cl}}\rangle =\left( \omega _{\mathrm{c}}+4\kappa \frac{G^2}{\omega _{\mathrm{m}}} \right) \alpha ^2-2G\sqrt{N_{\mathrm{B}}}\alpha \sin \theta +\frac{\omega _{\mathrm{m}}}{2}N_{\mathrm{B}}\left( 1 -\cos \theta \right) 
\end{equation}
We check the first-order partial derivatives and Hessian Matrix
\begin{gather}
\frac{\partial H_{\mathrm{cl}}^{\left( \kappa \right)}}{\partial \theta}=\frac{\omega _{\mathrm{m}}}{2}N_{\mathrm{B}}\sin \theta -\frac{G}{\sqrt{N_{\mathrm{B}}}}2\alpha N_{\mathrm{B}}\cos \theta =0
\\
\frac{\partial H_{\mathrm{cl}}^{\left( \kappa \right)}}{\partial \alpha}=2\left( \omega _{\mathrm{c}}+4\kappa \frac{G^2}{\omega _{\mathrm{m}}} \right) \alpha -2\frac{G}{\sqrt{N_{\mathrm{B}}}}N_{\mathrm{B}}\sin \theta =0
\\
\mathbb{H} =\left( \begin{matrix}
	\frac{\partial ^2H_{\mathrm{cl}}^{\left( \kappa \right)}}{\partial \alpha \partial \alpha}&		\frac{\partial ^2H_{\mathrm{cl}}^{\left( \kappa \right)}}{\partial \alpha \partial \theta}\\
	\frac{\partial ^2H_{\mathrm{cl}}^{\left( \kappa \right)}}{\partial \theta \partial \alpha}&		\frac{\partial ^2H_{\mathrm{cl}}^{\left( \kappa \right)}}{\partial \theta \partial \theta}\\
\end{matrix} \right) =\left( \begin{matrix}
	2\left( \omega _{\mathrm{c}}+4\kappa \frac{G^2}{\omega _{\mathrm{m}}} \right)&		-2G\sqrt{N_{\mathrm{B}}}\cos \theta\\
	-2G\sqrt{N_{\mathrm{B}}}\cos \theta&		\frac{\omega _{\mathrm{m}}}{2}N_{\mathrm{B}}\cos \theta +2\alpha G\sqrt{N_{\mathrm{B}}}\sin \theta\\
\end{matrix} \right) 
\\
\det \left( \mathbb{H} \right) =2\left( \omega _{\mathrm{c}}+4\kappa \frac{G^2}{\omega _{\mathrm{m}}} \right) \left( \frac{\omega _{\mathrm{m}}}{2}N_{\mathrm{B}}\cos \theta +2\alpha G\sqrt{N_{\mathrm{B}}}\sin \theta \right) -\left( 2G\sqrt{N_{\mathrm{B}}}\cos \theta \right) ^2>0
\end{gather}
and solve the equations to find the lowest potential reached at 
\begin{equation}
\alpha =\frac{\omega _{\mathrm{m}}}{4G}\sqrt{N_{\mathrm{B}}}\tan \theta , \qquad                  \theta =\begin{cases}
	0,&		\qquad G\leqslant G_{\mathrm{c}}\\
	\pm \arccos \left( \kappa +\frac{\omega _{\mathrm{c}}\omega _{\mathrm{m}}}{4G^2} \right),&		\qquad G>G_{\mathrm{c}}\\
\end{cases}
\label{eqS:stable}
\end{equation}
The change of $\theta$ and $\alpha$ from zero to finite is due to SRPT, where the ground state contains a macroscopic number of excitations, while the vacuum state becomes a saddle point. The quantum phase transition occurs in the ground state without any thermal fluctuation. The critical point $G_{\mathrm{c}}=\sqrt{\frac{\omega _{\mathrm{m}}\omega _{\mathrm{c}}}{4\left( 1-\kappa \right)}}$ is deduced by $\det \left( \mathbb{H} \right) =0$. Notice the spontaneous symmetry breaking in the superradiant phase, where there is a pair of minimal potential points.

We add quantum fluctuation at the classical lowest potential points as an infinitesimal perturbation, by displacing $a$ and $a^{\dag}$ with $D_{\mathrm{A}}\left( \alpha \right) =e^{\alpha \left( a^{\dagger}-a \right)}$ and rotating $\sigma^\mathrm{x},\sigma^\mathrm{z}$ with $R_{\mathrm{B}}\left( \theta ,\sigma ^{\mathrm{y}} \right) =e^{-i\frac{\theta}{2}\sigma ^{\mathrm{y}}}$ to the coordinates $\alpha$ and $\theta$ in Eq.~(\ref{eqS:stable}):
\begin{gather}
a^{\prime}=D_{\mathrm{A}}\left( \alpha \right) aD_{\mathrm{A}}^{\dagger}\left( \alpha \right) =a-\alpha 
\\
\sigma ^{\mathrm{x}^{\prime}}=R_{\mathrm{B}}\left( \theta ,\sigma ^{\mathrm{y}} \right) \sigma ^{\mathrm{x}}R_{\mathrm{B}}^{\dagger}\left( \theta ,\sigma ^{\mathrm{y}} \right) =\cos \theta \sigma ^{\mathrm{x}}-\sin \theta \sigma ^{\mathrm{z}}
\\
\sigma ^{\mathrm{z}^{\prime}}=R_{\mathrm{B}}\left( \theta ,\sigma ^{\mathrm{y}} \right) \sigma ^{\mathrm{z}}R_{\mathrm{B}}^{\dagger}\left( \theta ,\sigma ^{\mathrm{y}} \right) =\sin \theta \sigma ^{\mathrm{x}}+\cos \theta \sigma ^{\mathrm{z}}
\end{gather}
squeezed Dicke Hamiltonian can then be written in the shifted bases (labeled by ``prime'') as
\begin{equation}
\begin{aligned}
H^{\left( \kappa \right)}=
& 
\omega _{\mathrm{c}}a^{\prime\dagger}a^{\prime}+\kappa \frac{G^2}{\omega _{\mathrm{m}}}\left( a^{\prime\dagger}+a^{\prime} \right) ^2+\frac{\omega _{\mathrm{m}}}{\cos \theta}\left( \frac{N_{\mathrm{B}}}{2}+S^{\mathrm{z}\prime} \right) -\frac{\left( 1-\cos \theta \right) ^2}{2\cos \theta}N_{\mathrm{B}}\frac{\omega _{\mathrm{m}}}{2}
\\
&
+2\frac{G}{\sqrt{N_{\mathrm{B}}}}\left( a^{\prime\dagger}+a^{\prime} \right) \cos \theta S^{\mathrm{x}\prime}+\frac{2G\sin \theta}{\sqrt{N_{\mathrm{B}}}}\left( \frac{N_{\mathrm{B}}}{2}+S^{\mathrm{z}\prime} \right) \left( a^{\prime\dagger}+a^{\prime} \right) 
\end{aligned}
\label{eqS:ShiftedHamiltonian}
\end{equation}

Because the transition dipole between dark states and polaritons is zero because of different total angular momentum quantum numbers and dark states are usually of high energy state with no population for thermal state at room temperature, we are only interested in polariton states that contribute to the QFI. So we focus on $s=N_{\mathrm{B}}/2$ subspace and turn the Dicke operators into a bosonic operator in thermodynamics limit ($N_{\mathrm{B}}\rightarrow 0$) by Holstein–Primakoff transformation:
\begin{gather}
S^{+^{\prime}}=b^{\prime\dagger}\sqrt{2s-b^{\prime\dagger}b^{\prime}},\qquad 
S^{-^{\prime}}=\sqrt{2s-b^{\prime\dagger}b^{\prime}}b^{\prime},\qquad S^{\mathrm{z}^{\prime}}=b^{\prime\dagger}b^{\prime}-s
\label{eqS:HolsteinPrimakoffTransformation}
\end{gather}
In this definition, the spin-down ferromagnetic Dicke state $|\frac{N_{\mathrm{B}}}{2},-\frac{N_{\mathrm{B}}}{2}\rangle $ is the vacuum state. Since we have already shifted the Hamiltonian to the minimal potential point, we can treat quantum fluctuation infinitesimally as $a^{\prime\dagger}a^{\prime}, b^{\prime\dagger}b^{\prime}\ll N_{\mathrm{B}}$. So we can Tayler expend Eq.~(\ref{eqS:HolsteinPrimakoffTransformation}) as
 $S^{\mathrm{x}}=\sqrt{N_{\mathrm{B}}}\left( b^{\prime\dagger}+b^{\prime} \right)$ 
and drop cubic or higher terms in Eq.~(\ref{eqS:ShiftedHamiltonian})
\begin{equation}
H^{\left( \kappa \right)}=\omega _{\mathrm{c}}a^{\prime\dagger}a^{\prime}+\kappa \frac{G^2}{\omega _{\mathrm{m}}}\left( a^{\prime\dagger}+a^{\prime} \right) ^2+\frac{\omega _{\mathrm{m}}}{\cos \theta}b^{\prime\dagger}b^{\prime}-\frac{\left( 1-\cos \theta \right) ^2}{2\cos \theta}N_{\mathrm{B}}\frac{\omega _{\mathrm{m}}}{2}+G\cos \theta \left( a^{\prime\dagger}+a^{\prime} \right) \left( b^{\prime\dagger}+b^{\prime} \right)
\label{eqS:HPAHamiltonian}
\end{equation}

Now we have a quadratic bosonic Hamiltonian, which can be easily diagonalized by Bogoliubov Transformation
\begin{equation}
\left( \begin{array}{c}
	a^{\prime}\\
	b^{\prime}\\
	{a^{\prime}}^{\dagger}\\
	{b^{\prime}}^{\dagger}\\
\end{array} \right) =\left( \begin{matrix}
	u&		v\\
	v^*&		u^*\\
\end{matrix} \right) \left( \begin{array}{c}
	c_{+}^{\prime}\\
	c_{-}^{\prime}\\
	{c_{+}^{\prime}}^{\dagger}\\
	{c_{-}^{\prime}}^{\dagger}\\
\end{array} \right) 
\end{equation}
The symplectic matrix entries for Bogoliubov transformation are 
\begin{gather}
u=\left( \begin{matrix}
	\frac{1}{2}\cos \frac{\zeta}{2}\frac{\omega _{\mathrm{c}}+\Omega _{+}^{\prime}}{\sqrt{\omega _{\mathrm{c}}\Omega _{+}^{\prime}}}&		-\frac{1}{2}\sin \frac{\zeta}{2}\frac{\omega _{\mathrm{c}}+\Omega _{-}^{\prime}}{\sqrt{\omega _{\mathrm{c}}\Omega _{-}^{\prime}}}\\
	\frac{1}{2}\sin \frac{\zeta}{2}\frac{\frac{\omega _{\mathrm{m}}}{\cos \theta}+\Omega _{+}^{\prime}}{\sqrt{\frac{\omega _{\mathrm{m}}}{\cos \theta}\Omega _{+}^{\prime}}}&		\frac{1}{2}\cos \frac{\zeta}{2}\frac{\frac{\omega _{\mathrm{m}}}{\cos \theta}+\Omega _{-}^{\prime}}{\sqrt{\frac{\omega _{\mathrm{m}}}{\cos \theta}\Omega _{-}^{\prime}}}\\
\end{matrix} \right) ,
\qquad
v=\left( \begin{matrix}
	\frac{1}{2}\cos \frac{\zeta}{2}\frac{\omega _{\mathrm{c}}-\Omega _{+}^{\prime}}{\sqrt{\omega _{\mathrm{c}}\Omega _{+}^{\prime}}}&		-\frac{1}{2}\sin \frac{\zeta}{2}\frac{\omega _{\mathrm{c}}-\Omega _{-}^{\prime}}{\sqrt{\omega _{\mathrm{c}}\Omega _{-}^{\prime}}}\\
	\frac{1}{2}\sin \frac{\zeta}{2}\frac{\frac{\omega _{\mathrm{m}}}{\cos \theta}-\Omega _{+}^{\prime}}{\sqrt{\frac{\omega _{\mathrm{m}}}{\cos \theta}\Omega _{+}^{\prime}}}&		\frac{1}{2}\cos \frac{\zeta}{2}\frac{\frac{\omega _{\mathrm{m}}}{\cos \theta}-\Omega _{-}^{\prime}}{\sqrt{\frac{\omega _{\mathrm{m}}}{\cos \theta}\Omega _{-}^{\prime}}}\\
\end{matrix} \right) 
\end{gather}
where $\cos \zeta =\frac{\frac{1}{2}\left( \omega _{\mathrm{c}}^{2}+4\kappa \frac{G^2}{\omega _{\mathrm{m}}}\omega _{\mathrm{c}}-\frac{\omega _{\mathrm{m}}^{2}}{\cos ^2\theta} \right)}{\sqrt{\left( 2G\sqrt{\omega _{\mathrm{c}}\omega _{\mathrm{m}}\cos \theta} \right) ^2+\frac{1}{4}\left( \omega _{\mathrm{c}}^{2}+4\kappa \frac{G^2}{\omega _{\mathrm{m}}}\omega _{\mathrm{c}}-\frac{\omega _{\mathrm{m}}^{2}}{\cos ^2\theta} \right) ^2}}$. Bogoliubov transformation results in two independent bosonic modes, $\Omega _{+}^{\prime}$ and $\Omega _{-}^{\prime}$, that are commute to each other. These two quasi-particles are actually the upper polariton and lower polariton. The diagonalized Hamiltonian is 
\begin{equation}
    H^{\left( \kappa \right)}=\Omega _{+}^{\prime}{c_{+}^{\prime}}^{\dagger}c_{+}^{\prime}+\Omega _{-}^{\prime}{c_{-}^{\prime}}^{\dagger}c_{-}^{\prime}+E_{0_+,0_-}
\end{equation}
with the polariton modes and zero-point energy
\begin{gather}
\Omega _{\pm}^{\prime}=\sqrt{\frac{1}{2}\left( \omega _{\mathrm{c}}^{2}+4\kappa \frac{G^2}{\omega _{\mathrm{m}}}\omega _{\mathrm{c}}+\frac{\omega _{\mathrm{m}}^{2}}{\cos ^2\theta} \right) \pm \frac{1}{2}\sqrt{\left( \omega _{\mathrm{c}}^{2}+4\kappa \frac{G^2}{\omega _{\mathrm{m}}}\omega _{\mathrm{c}}-\frac{\omega _{\mathrm{m}}^{2}}{\cos ^2\theta} \right) ^2+16G^2\omega _{\mathrm{c}}\omega _{\mathrm{m}}\cos \theta}}
\\
E_{0_+,0_-}=\frac{1}{2}\Omega _{+}^{\prime}+\frac{1}{2}\Omega _{-}^{\prime}-\frac{1}{2}N_{\mathrm{B}}\omega _{\mathrm{m}}\frac{\left( 1-\cos \theta \right) ^2}{2\cos \theta}-\frac{1}{2}\omega _{\mathrm{c}}-\frac{1}{2}\frac{\omega _{\mathrm{m}}}{\cos \theta}
\end{gather}
In particular, under the rotating wave approximation $\left(\left| \omega _{\mathrm{c}}-\omega _{\mathrm{m}} \right| \sim G \ll \omega _{\mathrm{c}}+\omega _{\mathrm{m}}\right)$, the polariton modes become $\Omega _{\pm}^{\prime}\approx \omega _{\mathrm{m}}\pm G$, recovering the Tavis-Cummings model. The eigenstates are simply the tensor product of two independent Fock states $|n_+,n_-\rangle =|n_+\rangle \otimes |n_-\rangle $ and the eigenenergies are $E_{n_+,n_-}=\Omega _{+}^{\prime}n_++\Omega _{-}^{\prime}n_-+E_{0_+,0_-}$. The thermal state of the squeezed Dicke Model is simply the tensor product of the two polariton modes
\begin{equation}
\rho \left( T \right) =\left( 1-e^{-\beta \Omega _{+}^{\prime}} \right) \left( 1-e^{-\beta \Omega _{-}^{\prime}} \right) \left( \sum_{n_+}{e^{-\beta \Omega _{+}^{\prime}n_+}|n_+\rangle \langle n_+|} \right) \otimes \left( \sum_{n_-}{e^{-\beta \Omega _{-}^{\prime}n_-}|n_-\rangle \langle n_-|} \right) 
\label{eqS:thermalstate}
\end{equation}
and the Boltzmann distribution is
\begin{equation}
p_{n_+,n_-}=\left( 1-e^{-\beta \Omega _{+}^{\prime}} \right) \left( 1-e^{-\beta \Omega _{-}^{\prime}} \right) e^{-\beta \left( \Omega _{+}^{\prime}n_++\Omega _{-}^{\prime}n_- \right)}
\end{equation}
Note that the thermal state is built from the ground state quantum phase transition, which can not capture the quantum critical region, but this structure is still a good approximation when $\Omega_\pm \gg T$ for the electronic energy scale.

The spontaneous symmetry breaking occurs for the parity operator $\mathcal{P} =e^{i\pi \left( a^{\dagger}a+S^{\mathrm{z}} \right)}$, which rotates the Dicke state along the z-axis of the Bloch sphere by $\pi$ and rotates the photon state around the origin of the phase space by $\pi$. In the normal phase, the isotropic vacuum is the ground state and there is no special orientation, so the parity operator keeps the eigenstates unchanged $\mathcal{P} |n_+,n_-\rangle =|n_+,n_-\rangle $, or $\left[ H^{\left( \kappa \right)},\mathcal{P} \right] =0$. In superradiant phase, spontaneous symmetry breaking of parity makes the total Helbert space as a direct sum of two asymmetric subspaces $\left(\mathcal{H} =\mathcal{H} _{\mathrm{L}}\oplus \mathcal{H} _{\mathrm{R}}\right)$, and the total Hamiltonian gets decomposed as $H^{\left( \kappa \right)}=H_{\mathrm{L}}^{\left( \kappa \right)}\oplus H_{\mathrm{R}}^{\left( \kappa \right)}$, donated by L and R for the different sign of $\theta$ in Eq.~(\ref{eqS:stable}). However, Eq.~(\ref{eqS:HPAHamiltonian}) does not depend on the choice of the sign of $\theta$ because it only depends on $\cos\theta$. Thus the two subspaces are the same copy of each other, where the Hamiltonian is exactly the same and has the same energy spectrum. Thus each polariton mode is 2-fold degenerate above the phase transition. Treating the parity operator $\mathcal{P} $ as an operation of rotation, we can build a mapping between the two subspace via parity operator: $\mathcal{P} |n_+,n_-\rangle _{\mathrm{L}}=|n_+,n_-\rangle _{\mathrm{R}}$ and $\mathcal{P} |n_+,n_-\rangle _{\mathrm{R}}=|n_+,n_-\rangle _{\mathrm{L}}$, or $\mathcal{P} H_{\mathrm{L}}^{\left( \kappa \right)}\mathcal{P} ^{-1}=H_{\mathrm{R}}^{\left( \kappa \right)}$ and $\mathcal{P} H_{\mathrm{R}}^{\left( \kappa \right)}\mathcal{P} ^{-1}=H_{\mathrm{L}}^{\left( \kappa \right)}$. So it is clear that $\left[ H^{\left( \kappa \right)},\mathcal{P} \right] \ne 0$ and thus no global parity symmetry. Instead, local parity still conserved as $\mathcal{P} ^{\prime}=e^{i\pi \left( c_{+}^{\prime\dagger}c_{+}^{\prime}+c_{-}^{\prime\dagger}c_{-}^{\prime} \right)}$ in each subspace. Lower polariton mode $\Omega _{-}^{\prime}$ also works as the characteristic energy for SRPT that goes to zero at $G_{\mathrm{c}}$.

If we take the state to be equally-weighted direct sum $\rho  =\frac{1}{2}\rho _{\mathrm{L}} \oplus \frac{1}{2}\rho _{\mathrm{R}} $, we maintain the global parity symmetry $\mathcal{P}$. We do this also because it is hard to separate one subspace from another in the experiments. For example, the thermal state in superradiant phase is the equally weighted direct sum of the thermal states in the 2 subspaces due to symmetry breaking, $\rho \left( T \right) =\frac{1}{2}\rho _{\mathrm{L}}\left( T \right) \oplus \frac{1}{2}\rho _{\mathrm{R}}\left( T \right) $, while $\rho _{\mathrm{L}}\left( T \right)$ and $\rho _{\mathrm{R}}\left( T \right)$ are the same as Eq.~(\ref{eqS:thermalstate}).

\begin{figure}[htbp]
    \centering
    \includegraphics[width=1\textwidth]{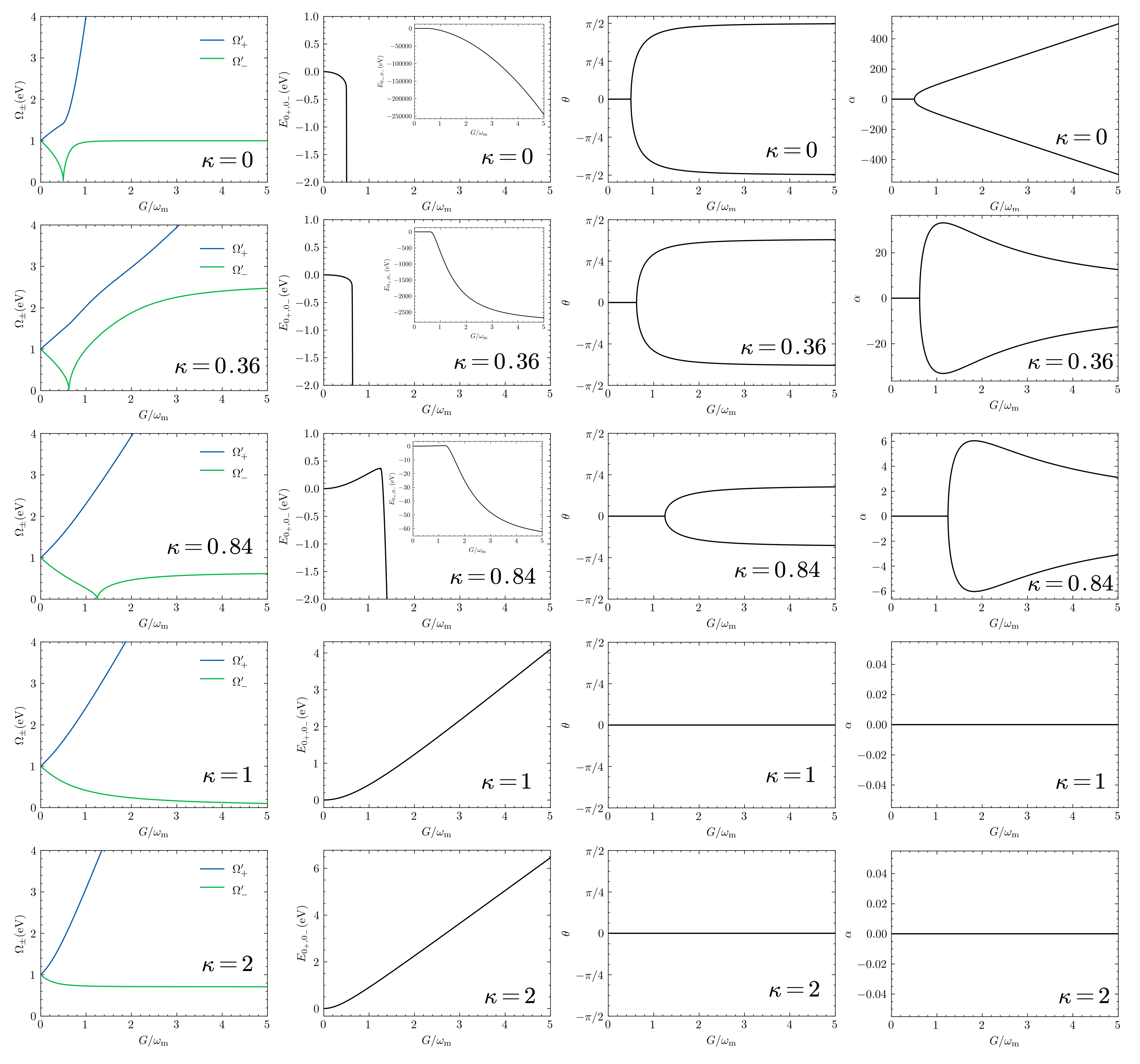}
    \caption{Analytical solution for $H^{\mathrm{\kappa}}$ as a function of $G$. From left to right: mode frequencies of upper polariton ($\Omega^{\prime}_{+}$) and lower polariton ($\Omega^{\prime}_{-}$); zero-point energy $E_{0_+,0_-}$ (insets are the same data in larger scale); Dicke state rotation angle $\theta$; photon state displacement $\alpha$.  From up to down: $\kappa=0$; $\kappa=0.36$; $\kappa=0.84$; $\kappa=1$; $\kappa=2$. Parameters: $\omega_{\mathrm{m}}=\omega_{\mathrm{c}}=1\si{eV}$. }
    \label{figS:Seigen}
\end{figure}

Some data are shown for the analytical solution for the squeezed Dicke model in Fig.~\ref{figS:Seigen}.

\section{Maximized QFI}
In this section, we maximize by choosing the local spin operator with the same bandwidth of 1 but in a different spin direction, so that QFI can be a better entanglement witness. Maximized QFI is defined as 
\begin{equation}
F_{\mathrm{Q}}^{\mathrm{Max}}\left[ \rho \right] =\underset{\vec{n}}{\max}\left\{ F_{\mathrm{Q}}\left[ \rho ,\vec{n}\cdot \vec{S} \right] \right\} 
\end{equation}
where $\vec{n}\cdot \vec{S}=\sum_{\gamma =\mathrm{x},\mathrm{y},\mathrm{z}}{n_{\gamma}}S^{\gamma}$ and $\vec{n}$ is a unit vector. Since $F_{\mathrm{Q}}\left[ \rho,\vec{n}\cdot \vec{S} \right] $ can be explicitly written as 
\begin{equation}
F_{\mathrm{Q}}\left[ \rho ,\vec{n}\cdot \vec{S} \right] =\sum_{\gamma _1,\gamma _2}{n_{\gamma _1}n_{\gamma _2}\left( \sum_{l,l^{\prime}}{2\frac{\left( p_l-p_{l^{\prime}} \right) ^2}{p_l+p_{l^{\prime}}}\mathrm{Re}\left( \langle l|S^{\gamma _1}|l^{\prime}\rangle \langle l^{\prime}|S^{\gamma _2}|l\rangle \right)} \right)}=\sum_{\gamma _1,\gamma _2}{n_{\gamma _1}n_{\gamma _2}\mathcal{F} ^{\gamma _1\gamma _2}}
\end{equation}
where $\mathcal{F} $ is a positive-semi definite matrix. So diagonalize $\mathcal{F} $ and the maximal eigenvalue of $\mathcal{F} $ will be $F_{\mathrm{Q}}^{\mathrm{Max}}\left[ \rho\right]$. 

Actually, due to parity symmetry, $F_{\mathrm{Q}}^{\mathrm{Max}}\left[ \rho \right]$ has an even simpler form. Bacause the parity operator $\mathcal{P} ^{\prime}=e^{i\pi \left( c_{+}^{\prime\dagger}c_{+}^{\prime}+c_{-}^{\prime\dagger}c_{-}^{\prime} \right)}$ commutes to $H^{\left(\kappa\right)}$ as $\left[ \mathcal{P} ^{\prime},H^{\left( \kappa \right)} \right] =0$, the eigenstate of $H^{\left(\kappa\right)}$ are of certain parity (either even or odd) with no uncertainty (zero-variance) and the parity is conserved during the time-evolution. Because $S^{\mathrm{z}}$ conserves the parity while $S^{\mathrm{x}}, S^{\mathrm{y}}$ change the parity, $|l\rangle $ and $|l^{\prime}\rangle $ needs to be of same parity to make $S^{\mathrm{z}}$ matrix elements non-zero, but same parity makes $S^{\mathrm{x}}, S^{\mathrm{y}}$ matrix elements non-zero. So we have
\begin{equation}
\mathcal{F} ^{\mathrm{zx}}=\mathcal{F} ^{\mathrm{xz}}=\mathcal{F} ^{\mathrm{yz}}=\mathcal{F} ^{\mathrm{zy}}=0
\end{equation}
As for $\mathcal{F} ^{\mathrm{yx}}=\mathcal{F} ^{\mathrm{xy}}$ terms, we need to write the operators in the shifted bases as $S^{\mathrm{x}}=\cos \theta S^{\mathrm{x}\prime}+\sin \theta S^{\mathrm{z}\prime}$ and $S^{\mathrm{y}}=S^{\mathrm{y}\prime}$ and do the Holstein-Primakoff Transformation: $2\mathrm{Re}\left( \langle l|S^{\mathrm{x}}|l^{\prime}\rangle \langle l^{\prime}|S^{\mathrm{y}}|l\rangle \right) =2\cos \theta \mathrm{Re}\left( \langle l|S^{\mathrm{x}\prime}|l^{\prime}\rangle \langle l^{\prime}|S^{\mathrm{y}\prime}|l\rangle \right) =\frac{N_{\mathrm{B}}}{2i}\cos \theta \left( \langle l|b^{\prime\dagger}|l^{\prime}\rangle \langle l^{\prime}|b^{\prime\dagger}|l\rangle -\langle l|b^{\prime}|l^{\prime}\rangle \langle l^{\prime}|b^{\prime}|l\rangle \right) $ and we translate rotated bosons into polariton bases as $\langle l|b^{\prime\dagger}|l^{\prime}\rangle \langle l^{\prime}|b^{\prime\dagger}|l\rangle =v_{21}u_{21}\langle l|c_{+}^{\prime}|l^{\prime}\rangle \langle l^{\prime}|{c_{+}^{\prime}}^{\dagger}|l\rangle +v_{22}u_{22}\langle l|c_{-}^{\prime}|l^{\prime}\rangle \langle l^{\prime}|{c_{-}^{\prime}}^{\dagger}|l\rangle +u_{21}v_{21}\langle l|{c_{+}^{\prime}}^{\dagger}|l^{\prime}\rangle \langle l^{\prime}|c_{+}^{\prime}|l\rangle +u_{22}v_{22}\langle l|{c_{-}^{\prime}}^{\dagger}|l^{\prime}\rangle \langle l^{\prime}|c_{-}^{\prime}|l\rangle $ which is real. So we have $\langle l|b^{\prime\dagger}|l^{\prime}\rangle \langle l^{\prime}|b^{\prime\dagger}|l\rangle =\langle l|b^{\prime}|l^{\prime}\rangle \langle l^{\prime}|b^{\prime}|l\rangle $ and 
\begin{equation}
\mathcal{F} ^{\mathrm{yx}}=\mathcal{F} ^{\mathrm{xy}}=0
\end{equation}
Therefore, $\mathcal{F}$ is already diagonalized and we only need to check the diagonal terms of it
\begin{equation}
f_{\mathrm{Q}}^{\mathrm{Max}}\left[ \rho \right] =\max \left\{ f_{\mathrm{Q}}\left[ \rho ,S^{\mathrm{x}} \right] , f_{\mathrm{Q}}\left[ \rho ,S^{\mathrm{y}} \right] , f_{\mathrm{Q}}\left[ \rho ,S^{\mathrm{z}} \right] \right\}
\label{eqS:maxQFI}
\end{equation}
Notice that even though we derive the Eq.~(\ref{eqS:maxQFI}) in the thermodynamic limit, this result is still valid for the finite-size system as long as the parity symmetry holds.

\section{Exact Diagonalization of QFI for finite size system}
Here we only focus on the numerical results for the thermal state via Exact Diagonalization. We show more data as supplements. For the thermal states for cavity QDE at electronic coupling regime, it is a good approximation to consider only the two lowest states with microscopy population (i.e, the ground state and the single lower polariton state) and the other high energy are all of zero occupation due to large energy gap. In the two-state approximation framework, the density matrix, purity, and QFI are
\begin{gather}
\rho \left( T \right) \approx \frac{1+\xi \left( T \right)}{2}|0\rangle \langle 0|+\frac{1-\xi \left( T \right)}{2}|1\rangle \langle 1|
\\
\gamma \left( T \right) \approx \left( \frac{1}{1+e^{-\left( \epsilon _1-\epsilon _0 \right) /k_{\mathrm{B}}T}} \right) ^2+\left( \frac{1}{1+e^{\left( \epsilon _1-\epsilon _0 \right) /k_{\mathrm{B}}T}} \right) ^2=\frac{1+\xi ^2\left( T \right)}{2}
\\
F_{\mathrm{Q}}\left[ \rho \left( T \right) ,O \right] \approx \frac{1+\xi \left( T \right)}{2}F_{\mathrm{Q}}\left[ |0\rangle ,O \right] +\frac{1-\xi \left( T \right)}{2}F_{\mathrm{Q}}\left[ |1\rangle ,O \right] -4\left( 1-\xi ^2\left( T \right) \right) \left| \langle 1|O|0\rangle \right|^2
\end{gather}
where $\xi \left( T \right) =\tanh \left( \frac{\epsilon _1-\epsilon _0}{2k_{\mathrm{B}}T} \right) $. And we have $F_{\mathrm{Q}}^{\mathrm{Max}}\left[ \rho \right] =\max \left\{ F_{\mathrm{Q}}\left[ \rho ,S^{\mathrm{x}} \right] , F_{\mathrm{Q}}\left[ \rho ,S^{\mathrm{y}} \right] , F_{\mathrm{Q}}\left[ \rho ,S^{\mathrm{z}} \right] \right\}$.

\begin{figure}[htbp]
    \centering
    \includegraphics[width=1\textwidth]{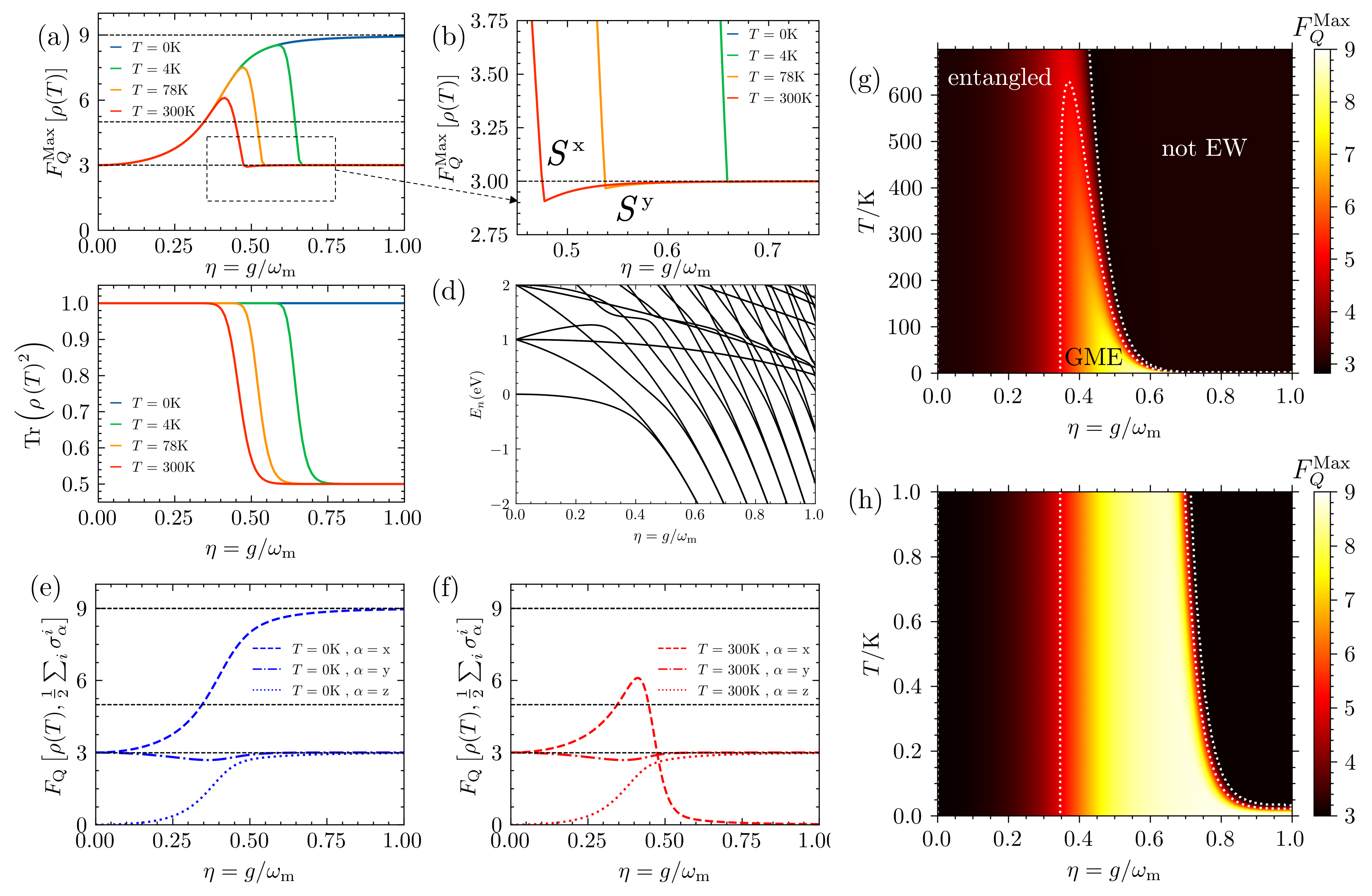}
    \caption{Numerical solution for Dicke model ($\kappa=0$ and $N_{\mathrm{B}}=3$) thermal state QFI. (a) Maximized QFI for the thermal state as a function of $G$ for $H^{\left( 0 \right)}$ and (b) a zoomed-in figure. The dashed horizontal lines are the thresholds $F\left(K\right)$ for EW. (c) Purity for the thermal state as a function of $G$. (d) Eigenspectrum $E_n$ as a function of $G$ for Dicke model $H^{\left( 0 \right)}$. (e)(f) a comparison between $ F_{\mathrm{Q}}\left[ \rho ,S^{\mathrm{x}} \right] , F_{\mathrm{Q}}\left[ \rho ,S^{\mathrm{y}} \right] , F_{\mathrm{Q}}\left[ \rho ,S^{\mathrm{z}} \right]$. (g) Phase diagram for QFI and (h) a zoomed-in figure. Notice that we describe the coupling strength by the interaction between a single molecule and photon $g=G/\sqrt{N_{\mathrm{B}}}$. Parameters: $\omega_{\mathrm{m}}=\omega_{\mathrm{c}}=1\si{eV}$. }
    \label{figS:Sthree}
\end{figure}

Consider three molecules ($N_{\mathrm{B}}$) coupled to the cavity. Dicke model $H^{\left( 0 \right)}$ eigenspectrum (Fig.~\ref{figS:Sthree}(d)) clearly shows the trend of two-fold degeneracy of single lower polariton and ground state at large $G$, which is a character of quantum phase transition in thermodynamics limit. Fig.~\ref{figS:Sthree}(a) shows that QFI is a good EW for zero temperature ground state ($F_{\mathrm{Q}}>F\left(1\right)=3$), and even becomes a molecular GME witness in the deep-strong coupling regime($F_{\mathrm{Q}}>F\left(2\right)=5$), asymptotically reaching the upper bound $F\left(3\right)=9$. At finite temperature, the $F_{\mathrm{Q}}^{\mathrm{Max}}$ increases as $G=g\sqrt{N_{\mathrm{B}}}$ increases by the same trend of $0\si{K}$, but then drops dramatically, because there is a macroscopic population on the first excited state (single lower polariton) due to degeneracy. In general, a more mixed state (less purity) means less QFI. Similarly, increasing temperature lowers the peaks in Fig.~\ref{figS:Sthree}(a). So for the Dicke model, QFI can work as a good EW in the ultrastrong and weak coupling regime near room temperature and may work well in the deep-strong coupling regime at lower temperatures. Notice that there is a sharp change when the maximized QFI (Fig.~\ref{figS:Sthree}(b)) is below $F\left(1\right)=3$ because the change of the optimized response operator from $S^{\mathrm{x}}$ to $S^{\mathrm{y}}$. However, QFI is not a good EW in this regime and it is of less interest to study $F_{\mathrm{Q}}\left[ \rho \left( T \right) ,S^{\mathrm{y}} \right] $. We can clearly see from Fig.~\ref{figS:Sthree}(e)(f) that in the regime of interest (entanglement witness works), $F_{\mathrm{Q}}\left[ \rho ,S^{\mathrm{x}} \right]$ maximizes QFI. We may also make a 2D figure of QFI as a function of $G$ and $T$ in the density profile (Fig.~\ref{figS:Sthree}(g)(h)) with the boundary for the three regimes (not EW, entangled, GME).

\begin{figure}[htbp]
    \centering
    \includegraphics[width=0.9\textwidth]{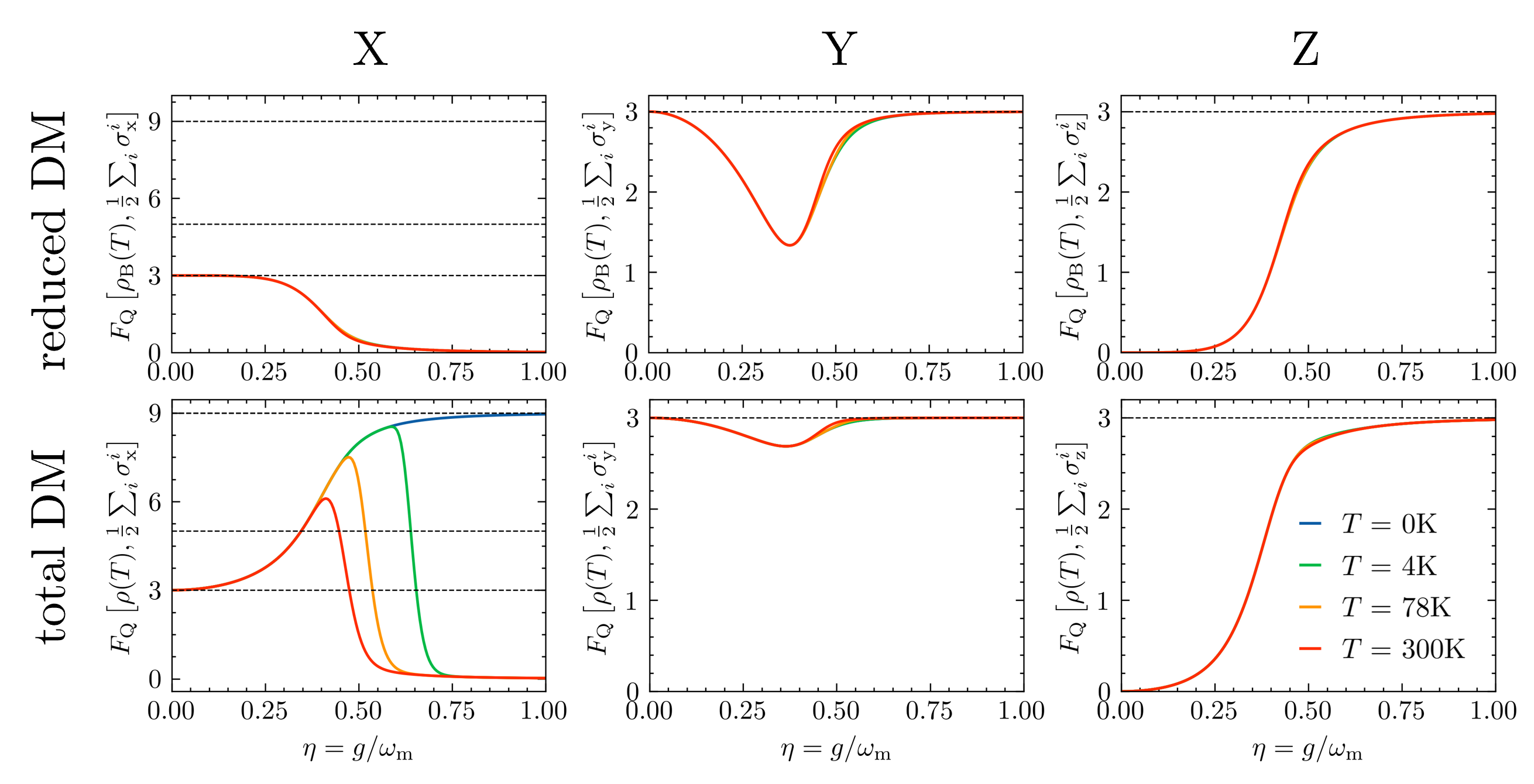}
    \caption{A comparison between QFI for reduced DM and QFI for total DM. From up to down: reduced DM and total DM. From left to right: $F_{\mathrm{Q}}\left[ \rho ,S^{\mathrm{x}} \right]$; $F_{\mathrm{Q}}\left[ \rho ,S^{\mathrm{y}} \right]$; $F_{\mathrm{Q}}\left[ \rho ,S^{\mathrm{z}} \right]$. Parameters: $\omega_{\mathrm{m}}=\omega_{\mathrm{c}}=1\si{eV}$, $\kappa=0$ and $N_{\mathrm{B}}=3$. }
    \label{figS:SreducedDM}
\end{figure}

We also want to show that when calculating QFI, we use the total density matrix (total DM, the state of both photons and molecules) rather than the reduced density matrix (reduced DM, the state of molecules after tracing our photon), because we are interesting in the entanglement. Theoretically, when studying multipartite entanglement within the subsystem of interest, of a total system, we need to look at the total density matrix rather than the reduced density, because tracing out the subsystem of no interest from the total system, the entanglement structure of the subsystem of interest change. An example is the GHZ state, which is a GME state $|\mathrm{GHZ}\rangle =\frac{1}{\sqrt{2}}\left( |111\rangle +|000\rangle \right) _{\mathrm{ABC}}$. However, if we trace out one of the particles, the other two particles become disentangled: $\mathrm{Tr}_{\mathrm{C}}\left( |\mathrm{GHZ}\rangle \langle \mathrm{GHZ}| \right) =\frac{1}{2}\left( |11\rangle \langle 11|+|00\rangle \langle 00| \right) _{\mathrm{AB}}$. For the same reason, as shown in Fig.~\ref{figS:SreducedDM}, QFI of total DM can detect molecular entanglement, but QFI of reduced DM fails. So we should not trace out the photon subsystem from the total system to study molecular entanglement. It also raises another problem with QFI measurement: the local measurement should be invariant between total DM and reduced DM, and thus total DM and reduced DM should give the same QFI. Actually, this statement is wrong, because QFI is dependent on the eigenstructure of the density matrix, and tracing out one of the subsystems changes the eigenstructure of the other subsystem. Therefore, QFI, in general, is not observable, and can only be measured in some special cases, such as the thermal state when given the information of dynamics or the pure states, as shown in the main text.

\begin{figure}[htbp]
    \centering
    \includegraphics[width=0.7\textwidth]{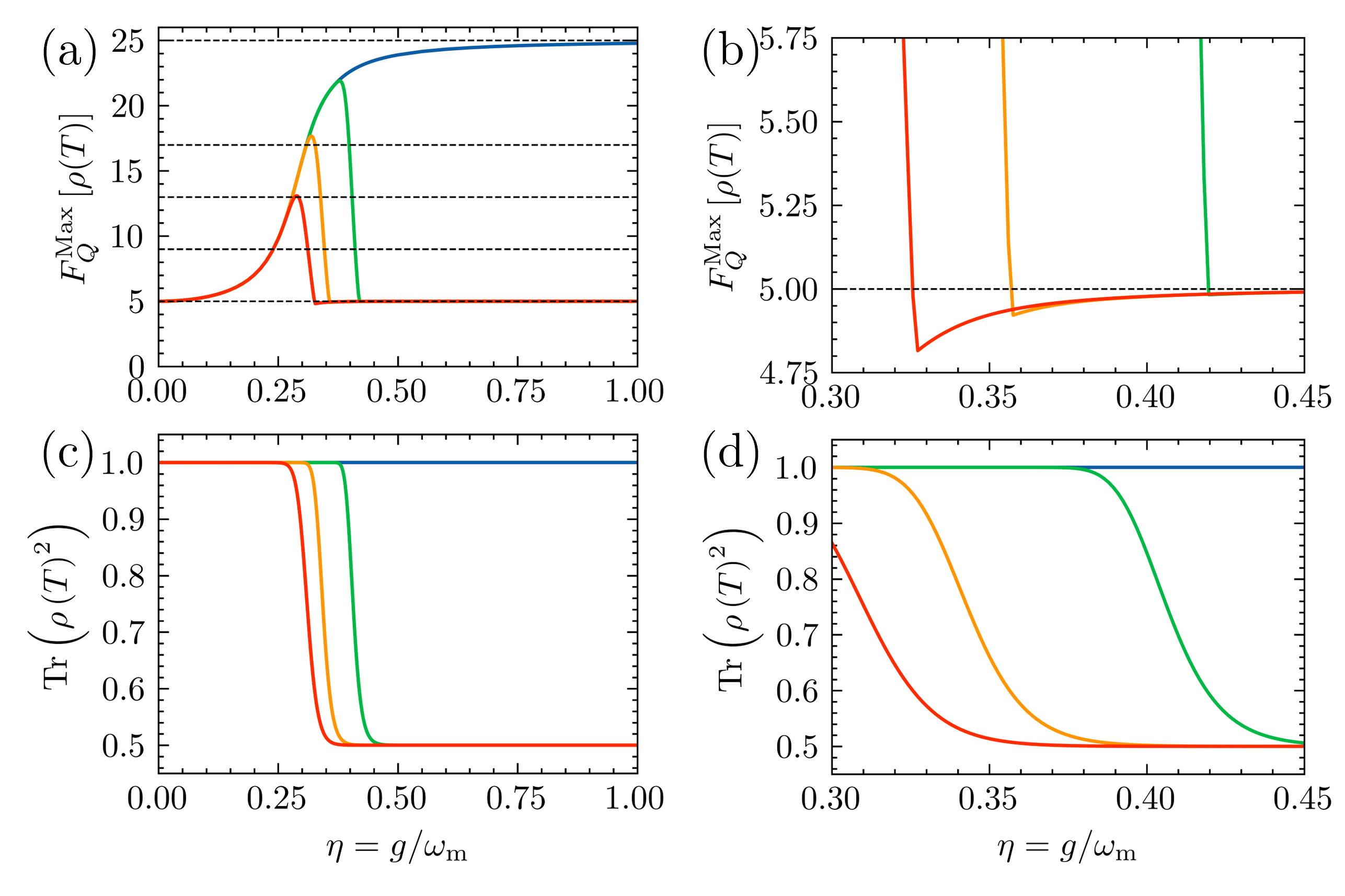}
    \caption{Numerical solution for Dicke model ($\kappa=0$ and $N_{\mathrm{B}}=5$) thermal state QFI. (a) Maximized QFI for the thermal state as a function of $G$ for $H^{\left( 0 \right)}$ and (b) a zoomed-in figure. The dashed horizontal lines are the thresholds $F\left(K\right)$ for EW. (c) Purity for the thermal state as a function of $G$ and (d) a zoomed-in figure. Parameters: $\omega_{\mathrm{m}}=\omega_{\mathrm{c}}=1\si{eV}$. }
    \label{figS:Sfive}
\end{figure}

We also give the data for $N_{\mathrm{B}}=5$, which shows a similar pattern (Fig.~\ref{figS:Sfive}).

\section{Analytical Expression of Thermal State QFI at thermodynamic limit}
In this section, we calculate QFI regarding the Dicke operators in three directions and get the analytical expression of maximized QFI. The eigenstates are written in the polariton Fock bases as $|n_+,n_-\rangle$ and the matrix elements are 
\begin{align}
    \langle m_+,m_-|S^{\mathrm{x}}|n_+,n_-\rangle 
    &=
    \cos \theta \langle m_+,m_-|S^{\mathrm{x}\prime}|n_+,n_-\rangle +\sin \theta \langle m_+,m_-|S^{\mathrm{z}\prime}|n_+,n_-\rangle 
    \\
    \langle m_+,m_-|S^{\mathrm{y}}|n_+,n_-\rangle
    &=
    \langle m_+,m_-|S^{\mathrm{y}\prime}|n_+,n_-\rangle 
    \\
    \langle m_+,m_-|S^{\mathrm{z}}|n_+,n_-\rangle 
    &
    =-\sin \theta \langle m_+,m_-|S^{\mathrm{x}\prime}|n_+,n_-\rangle +\cos \theta \langle m_+,m_-|S^{\mathrm{z}\prime}|n_+,n_-\rangle 
    \end{align}
and the norm square
\begin{align}
    \left| \langle m_+,m_-|S^{\mathrm{x}}|n_+,n_-\rangle \right|^2 
    &=
    \cos ^2\theta \left| \langle m_+,m_-|S^{\mathrm{x}\prime}|n_+,n_-\rangle \right|^2+\sin ^2\theta \left| \langle m_+,m_-|S^{\mathrm{z}\prime}|n_+,n_-\rangle \right|^2
    \\
    \left| \langle m_+,m_-|S^{\mathrm{y}}|n_+,n_-\rangle \right|^2 
    &=
    \left| \langle m_+,m_-|S^{\mathrm{y}\prime}|n_+,n_-\rangle \right|^2
    \\
    \left| \langle m_+,m_-|S^{\mathrm{z}}|n_+,n_-\rangle \right|^2 
    &=
    \sin ^2\theta \left| \langle m_+,m_-|S^{\mathrm{x}\prime}|n_+,n_-\rangle \right|^2+\cos ^2\theta \left| \langle m_+,m_-|S^{\mathrm{z}\prime}|n_+,n_-\rangle \right|^2
\end{align}
Notice that the crossing terms are eliminated due to the parity difference. Thus we have the sum rule:
\begin{align}
F_{\mathrm{Q}}\left[ \rho ,S^{\mathrm{x}} \right] 
&=
\cos ^2\theta F_{\mathrm{Q}}\left[ \rho ,S^{\mathrm{x}\prime} \right] +\sin ^2\theta F_{\mathrm{Q}}\left[ \rho ,S^{\mathrm{z}\prime} \right] 
\\
F_{\mathrm{Q}}\left[ \rho ,S^{\mathrm{y}} \right] 
&=
F_{\mathrm{Q}}\left[ \rho ,S^{\mathrm{y}\prime} \right] 
\\
F_{\mathrm{Q}}\left[ \rho ,S^{\mathrm{z}} \right] 
&=
\sin ^2\theta F_{\mathrm{Q}}\left[ \rho ,S^{\mathrm{x}\prime} \right] +\cos ^2\theta F_{\mathrm{Q}}\left[ \rho ,S^{\mathrm{z}\prime} \right] 
\end{align}
So we only need to calculate QFI regarding rotated Dicke operators. It is easier to use the asymmetric form of QFI to do the calculation $F_{\mathrm{Q}}\left[ \rho ,O \right]=4\sum_{l,{l^{\prime}}}^{\mathrm{dim}\left( \rho \right)}{p_l\frac{p_l-p_{l^{\prime}}}{p_l+p_{l^{\prime}}}\left| \langle l|O|{l^{\prime}}\rangle \right|^2}$.

For $x^{\prime}$ direction, we have 
\begin{equation}
\begin{aligned}
\left| \langle m_+,m_-|S^{\mathrm{x}\prime}|n_+,n_-\rangle \right|^2
= & 
\frac{N_{\mathrm{B}}}{4}\left( u_{21}+v_{21} \right) ^2\left( n_+\delta _{m_+,n_+-1}\delta _{m_-,n_-}+\left( n_++1 \right) \delta _{m_+,n_++1}\delta _{m_-,n_-} \right) 
\\
& +\frac{N_{\mathrm{B}}}{4}\left( u_{22}+v_{22} \right) ^2\left( n_-\delta _{m_+,n_+}\delta _{m_-,n_--1}+\left( n_-+1 \right) \delta _{m_+,n_+}\delta _{m_-,n_-+1} \right) 
\end{aligned}
\label{eqS:SxprimeMatrixSquare}
\end{equation}
\begin{equation}
\begin{aligned}
F_{\mathrm{Q}}\left[ \rho\left(T\right) ,S^{\mathrm{x}\prime} \right] 
=&
4\sum_{n_+,n_-}^{\infty}{p_{n_+,n_-}\sum_{m_+,m_-}^{\infty}{\frac{e^{-\beta \left( \Omega _{+}^{\prime}\left( n_+-m_+ \right) +\Omega _{-}^{\prime}\left( n_--m_- \right) \right)}-1}{e^{-\beta \left( \Omega _{+}^{\prime}\left( n_+-m_+ \right) +\Omega _{-}^{\prime}\left( n_--m_- \right) \right)}+1}\left| \langle m_+,m_-|S^{\mathrm{x}\prime}|n_+,n_-\rangle \right|^2}}
\\
=&
N_{\mathrm{B}}\left( \left( v_{21}+u_{21} \right) ^2\frac{e^{\beta \Omega _{+}^{\prime}}-1}{e^{\beta \Omega _{+}^{\prime}}+1}+\left( v_{22}+u_{22} \right) ^2\frac{e^{\beta \Omega _{-}^{\prime}}-1}{e^{\beta \Omega _{-}^{\prime}}+1} \right) 
\\
=&
N_{\mathrm{B}}\left( \sin ^2\frac{\zeta}{2}\frac{\frac{\omega _{\mathrm{m}}}{\cos \theta}}{\Omega _{+}^{\prime}}\tanh \left( \frac{\beta \Omega _{+}^{\prime}}{2} \right) +\cos ^2\frac{\zeta}{2}\frac{\frac{\omega _{\mathrm{m}}}{\cos \theta}}{\Omega _{-}^{\prime}}\tanh \left( \frac{\beta \Omega _{-}^{\prime}}{2} \right) \right) 
\end{aligned}
\end{equation}

For $y^{\prime}$ direction, we have 
\begin{equation}
\begin{aligned}
\left| \langle m_+,m_-|S^{\mathrm{y}\prime}|n_+,n_-\rangle \right|^2
=&
\frac{N_{\mathrm{B}}}{4}\left( u_{21}-v_{21} \right) ^2\left( \left( n_++1 \right) \delta _{m_+,n_++1}\delta _{m_-,n_-}+n_+\delta _{m_+,n_+-1}\delta _{m_-,n_-} \right) 
\\
&+\frac{N_{\mathrm{B}}}{4}\left( u_{22}-v_{22} \right) ^2\left( \left( n_-+1 \right) \delta _{m_+,n_+}\delta _{m_-,n_-+1}+n_-\delta _{m_+,n_+}\delta _{m_-,n_--1} \right) 
\end{aligned}
\label{eqS:SyprimeMatrixSquare}
\end{equation}
\begin{equation}
\begin{aligned}
F_{\mathrm{Q}}\left[ \rho\left(T\right) ,S^{\mathrm{y}\prime} \right] 
=&
4\sum_{n_+,n_-}^{\infty}{p_{n_+,n_-}\sum_{m_+,m_-}^{\infty}{\frac{e^{-\beta \left( \Omega _{+}^{\prime}\left( n_+-m_+ \right) +\Omega _{-}^{\prime}\left( n_--m_- \right) \right)}-1}{e^{-\beta \left( \Omega _{+}^{\prime}\left( n_+-m_+ \right) +\Omega _{-}^{\prime}\left( n_--m_- \right) \right)}+1}\left| \langle m_+,m_-|S^{\mathrm{y}\prime}|n_+,n_-\rangle \right|^2}}
\\
=&
N_{\mathrm{B}}\left( \left( u_{21}-v_{21} \right) ^2\frac{e^{\beta \Omega _{+}^{\prime}}-1}{e^{\beta \Omega _{+}^{\prime}}+1}+\left( u_{22}-v_{22} \right) ^2\frac{e^{\beta \Omega _{-}^{\prime}}-1}{e^{\beta \Omega _{-}^{\prime}}+1} \right) 
\\
=&
N_{\mathrm{B}}\left( \sin ^2\frac{\zeta}{2}\frac{\Omega _{+}^{\prime}}{\frac{\omega _{\mathrm{m}}}{\cos \theta}}\tanh \left( \frac{\beta \Omega _{+}^{\prime}}{2} \right) +\cos ^2\frac{\zeta}{2}\frac{\Omega _{-}^{\prime}}{\frac{\omega _{\mathrm{m}}}{\cos \theta}}\tanh \left( \frac{\beta \Omega _{-}^{\prime}}{2} \right) \right) 
\end{aligned}
\end{equation}

For $z^{\prime}$ direction, we have 
\begin{equation}
\begin{aligned}
    &\left| \langle m_+,m_-|S^{\mathrm{z}\prime}|n_+,n_-\rangle \right|^2
    \\
    =&\left( v_{21}u_{21} \right) ^2\left( n_+\left( n_+-1 \right) \delta _{m_+,n_+-2}\delta _{m_-,n_-}+\left( n_++2 \right) \left( n_++1 \right) \delta _{m_+,n_++2}\delta _{m_-,n_-} \right) 
    \\
    &+\left( u_{22}v_{22} \right) ^2\left( \left( n_-+1 \right) \left( n_-+2 \right) \delta _{m_+,n_+}\delta _{m_-,n_-+2}+n_-\left( n_--1 \right) \delta _{m_+,n_+}\delta _{m_-,n_--2} \right) 
    \\
    &+\left( u_{21}u_{22}+v_{22}v_{21} \right) ^2\left( \left( n_++1 \right) n_-\delta _{m_+,n_++1}\delta _{m_-,n_--1}+n_+\left( n_-+1 \right) \delta _{m_+,n_+-1}\delta _{m_-,n_-+1} \right) 
    \\
    &+\left( u_{21}v_{22}+u_{22}v_{21} \right) ^2\left( n_+n_-\delta _{m_+,n_+-1}\delta _{m_-,n_--1}+\left( n_++1 \right) \left( n_-+1 \right) \delta _{m_+,n_++1}\delta _{m_-,n_-+1} \right) 
    \\
    &+\left( \left( v_{22}v_{22}+u_{22}u_{22} \right) n_-+v_{22}v_{22}+\left( u_{21}u_{21}+v_{21}v_{21} \right) n_++v_{21}v_{21}-\frac{N_{\mathrm{B}}}{2} \right) ^2\delta _{m_+,n_+}\delta _{m_-,n_-}
\end{aligned}
\label{eqS:SzprimeMatrixSquare}
\end{equation}
\begin{equation}
\begin{aligned}
    F_{\mathrm{Q}}\left[ \rho\left(T\right) ,S^{\mathrm{z}\prime} \right] 
    =&
    4\sum_{n_+,n_-}^{\infty}{p_{n_+,n_-}\sum_{m_+,m_-}^{\infty}{\frac{e^{-\beta \left( \Omega _{+}^{\prime}\left( n_+-m_+ \right) +\Omega _{-}^{\prime}\left( n_--m_- \right) \right)}-1}{e^{-\beta \left( \Omega _{+}^{\prime}\left( n_+-m_+ \right) +\Omega _{-}^{\prime}\left( n_--m_- \right) \right)}+1}\left| \langle m_+,m_-|S^{\mathrm{z}\prime}|n_+,n_-\rangle \right|^2}}
    \\
    =&
    4\left( 1-e^{-\beta \Omega _{+}^{\prime}} \right) \left( 1-e^{-\beta \Omega _{-}^{\prime}} \right) \sum_{n_+,n_-}^{\infty}{e^{-\beta \left( \Omega _{+}^{\prime}n_++\Omega _{-}^{\prime}n_- \right)}\left( n_+\varXi _++n_-\varXi _-+\varXi _0 \right)}
    \\
    =&
    4\left( \frac{\varXi _+}{e^{\beta \Omega _{+}^{\prime}}-1}+\frac{\varXi _-}{e^{\beta \Omega _{-}^{\prime}}-1}+\varXi _0 \right) 
\end{aligned}
\end{equation}
where the parameters are given by
\begin{align}
\varXi _+
=&
4\left( v_{21}u_{21} \right) ^2\frac{e^{\beta \left( \Omega _{+}^{\prime}2 \right)}-1}{e^{\beta \left( \Omega _{+}^{\prime}2 \right)}+1}+\left( u_{21}v_{22}+u_{22}v_{21} \right) ^2\frac{e^{\beta \left( \Omega _{+}^{\prime}+\Omega _{-}^{\prime} \right)}-1}{e^{\beta \left( \Omega _{+}^{\prime}+\Omega _{-}^{\prime} \right)}+1}-\left( u_{21}u_{22}+v_{22}v_{21} \right) ^2\frac{e^{\beta \left( \Omega _{+}^{\prime}-\Omega _{-}^{\prime} \right)}-1}{e^{\beta \left( \Omega _{+}^{\prime}-\Omega _{-}^{\prime} \right)}+1}
\\
\varXi _-
=&
4\left( u_{22}v_{22} \right) ^2\frac{e^{\beta \left( \Omega _{-}^{\prime}2 \right)}-1}{e^{\beta \left( \Omega _{-}^{\prime}2 \right)}+1}+\left( u_{21}v_{22}+u_{22}v_{21} \right) ^2\frac{e^{\beta \left( \Omega _{+}^{\prime}+\Omega _{-}^{\prime} \right)}-1}{e^{\beta \left( \Omega _{+}^{\prime}+\Omega _{-}^{\prime} \right)}+1}+\left( u_{21}u_{22}+v_{22}v_{21} \right) ^2\frac{e^{\beta \left( \Omega _{+}^{\prime}-\Omega _{-}^{\prime} \right)}-1}{e^{\beta \left( \Omega _{+}^{\prime}-\Omega _{-}^{\prime} \right)}+1}
\\
\varXi _0
=&
\left( u_{21}v_{22}+u_{22}v_{21} \right) ^2\frac{e^{\beta \left( \Omega _{+}^{\prime}+\Omega _{-}^{\prime} \right)}-1}{e^{\beta \left( \Omega _{+}^{\prime}+\Omega _{-}^{\prime} \right)}+1}+2\left( v_{21}u_{21} \right) ^2\frac{e^{\beta \left( \Omega _{+}^{\prime}2 \right)}-1}{e^{\beta \left( \Omega _{+}^{\prime}2 \right)}+1}+2\left( u_{22}v_{22} \right) ^2\frac{e^{\beta \left( \Omega _{-}^{\prime}2 \right)}-1}{e^{\beta \left( \Omega _{-}^{\prime}2 \right)}+1}
\end{align}

\begin{gather}
\left( v_{21}u_{21} \right) ^2=\left( \frac{1}{4}\sin ^2\frac{\zeta}{2}\frac{\left( \frac{\omega _{\mathrm{m}}}{\cos \theta} \right) ^2-\left( \Omega _{+}^{\prime} \right) ^2}{\frac{\omega _{\mathrm{m}}}{\cos \theta}\Omega _{+}^{\prime}} \right) ^2
\\
\left( u_{22}v_{22} \right) ^2=\left( \frac{1}{4}\cos ^2\frac{\zeta}{2}\frac{\left( \frac{\omega _{\mathrm{m}}}{\cos \theta} \right) ^2-\left( \Omega _{-}^{\prime} \right) ^2}{\frac{\omega _{\mathrm{m}}}{\cos \theta}\Omega _{-}^{\prime}} \right) ^2
\\
\left( u_{21}v_{22}+u_{22}v_{21} \right) ^2=\left( \frac{1}{4}\sin \frac{\zeta}{2}\cos \frac{\zeta}{2}\frac{2\left( \frac{\omega _{\mathrm{m}}}{\cos \theta} \right) ^2-2\Omega _{+}^{\prime}\Omega _{-}^{\prime}}{\frac{\omega _{\mathrm{m}}}{\cos \theta}\sqrt{\Omega _{+}^{\prime}\Omega _{-}^{\prime}}} \right) ^2
\\
\left( u_{21}u_{22}+v_{22}v_{21} \right) ^2=\left( \frac{1}{4}\sin \frac{\zeta}{2}\cos \frac{\zeta}{2}\frac{2\left( \frac{\omega _{\mathrm{m}}}{\cos \theta} \right) ^2+2\Omega _{+}^{\prime}\Omega _{-}^{\prime}}{\frac{\omega _{\mathrm{m}}}{\cos \theta}\sqrt{\Omega _{+}^{\prime}\Omega _{-}^{\prime}}} \right) ^2
\end{gather}

Notice that both $F_{\mathrm{Q}}\left[ \rho\left(T\right) ,S^{\mathrm{x}\prime} \right] $ and $F_{\mathrm{Q}}\left[ \rho\left(T\right) ,S^{\mathrm{y}\prime} \right] $ are in the order of $O\left( N_{\mathrm{B}} \right)$ while $F_{\mathrm{Q}}\left[ \rho\left(T\right) ,S^{\mathrm{z}\prime} \right] $ is in the order of $O\left(1\right)$. Thus when calculating QFI per molecule $f_{\mathrm{Q}}\left[ \rho ,O \right] =F_{\mathrm{Q}}\left[ \rho ,O \right] /N_{\mathrm{B}}$, the contribution from $f_{\mathrm{Q}}\left[ \rho ,S^{\mathrm{z}\prime} \right] $ is negligible. So we have
\begin{align}
f_{\mathrm{Q}}\left[ \rho ,S^{\mathrm{x}} \right] 
=&
\cos ^2\theta f_{\mathrm{Q}}\left[ \rho ,S^{\mathrm{x}\prime} \right] 
\\
f_{\mathrm{Q}}\left[ \rho ,S^{\mathrm{y}} \right] 
=&
f_{\mathrm{Q}}\left[ \rho ,S^{\mathrm{y}\prime} \right] 
\\
f_{\mathrm{Q}}\left[ \rho ,S^{\mathrm{z}} \right] 
=&
\sin ^2\theta f_{\mathrm{Q}}\left[ \rho ,S^{\mathrm{x}\prime} \right] 
\end{align}

For the electronic coupling, we can simply ignore the population contribution from the upper polariton mode $\Omega _{+}^{\prime}$, while lower polariton mode $\Omega _{-}^{\prime}$ is v-shape regarding $G$. So approximately, we have $f_{\mathrm{Q}}\left[ \rho ,S^{\mathrm{x}} \right] \sim \cos ^2\frac{\zeta}{2}\cos \theta \frac{\omega _{\mathrm{m}}}{\Omega _{-}^{\prime}}$, which is $\lambda$-shape that can be larger than 1; $f_{\mathrm{Q}}\left[ \rho ,S^{\mathrm{y}} \right] \sim \cos ^2\frac{\zeta}{2}\frac{\Omega _{-}^{\prime}}{\omega _{\mathrm{m}}}$, which is v-shape that cannot be larger than 1; $f_{\mathrm{Q}}\left[ \rho ,S^{\mathrm{z}} \right] \sim \cos ^2\frac{\zeta}{2}\sin \theta \tan \theta \frac{\omega _{\mathrm{m}}}{\Omega _{-}^{\prime}}$, which is zero in the normal regime and monotonic increasing in the superradiant regime but cannot go beyond 1. Thus only $f_{\mathrm{Q}}\left[ \rho ,S^{\mathrm{x}} \right]$ is of interest, which can be a good entanglement witness. If $f_{\mathrm{Q}}\left[ \rho ,S^{\mathrm{x}} \right]<1$, maximized QFI fails to be a good entanglement witness. A more rigorous argument can be made by checking derivative $\frac{\partial}{\partial G}f_{\mathrm{Q}}\left[ \rho ,S^{\gamma} \right] $ and extreme values $f_{\mathrm{Q}}\left[ \rho ,S^{\gamma} \right] \mid_{G\rightarrow 0,G_{\mathrm{c}},\infty}^{}$.

QFI has the limits
\begin{align}
    & f_{\mathrm{Q}}\left[ \rho \left( T=0\mathrm{K} \right) ,S^{\mathrm{x}} \right] \mid _{G\rightarrow 0}^{}=1+O\left( G^2 \right) 
    \\
    &f_{\mathrm{Q}}\left[ \rho \left( T=0\mathrm{K} \right) ,S^{\mathrm{x}} \right] \mid _{G\rightarrow \infty}^{}=
    \begin{cases}
    	\frac{\kappa ^2}{\sqrt{1-\kappa ^2}},&\quad \kappa <1 \\
    	\frac{1}{\sqrt{1-\kappa ^{-1}}},&\quad \kappa \ge 1 \\
    \end{cases}
\label{eq:QFIlimit}
\end{align}
In the normal phase, increasing $\kappa$ reduces QFI, because external photon squeezing keeps the ground state vacuum and thus suppresses Dicke state squeezing.

\section{Analytical Expression of Eigenstate QFI at thermodynamic limit}
The eigenstate QFI can be similarly calculated. Consider the fact that QFI reduced to variance for the pure states:
\begin{equation}
F_{\mathrm{Q}}\left[ |\psi \rangle ,O \right] =4\langle \psi |O^2|\psi \rangle -4\langle \psi |O|\psi \rangle ^2
\end{equation}
So we only need to calculate $\langle \psi |O^2|\psi \rangle =\sum_{m_+,m_-}{\left| \langle m_+,m_-|O|\psi \rangle \right|}^2$ and $
\left| \langle \psi |O|\psi \rangle \right|^2$. Using Eq.~(\ref{eqS:SxprimeMatrixSquare}), Eq.~(\ref{eqS:SyprimeMatrixSquare}) and Eq.~(\ref{eqS:SzprimeMatrixSquare}), we can work out that

\begin{equation}
f_{\mathrm{Q}}\left[ |n_+,n_-\rangle ,S^{\mathrm{x}\prime} \right] = \sin ^2\frac{\zeta}{2}\frac{\frac{\omega _{\mathrm{m}}}{\cos \theta}}{\Omega _{+}^{\prime}}\left( 2n_++1 \right) +\cos ^2\frac{\zeta}{2}\frac{\frac{\omega _{\mathrm{m}}}{\cos \theta}}{\Omega _{-}^{\prime}}\left( 2n_-+1 \right) 
\end{equation}
\begin{equation}
f_{\mathrm{Q}}\left[ |n_+,n_-\rangle ,S^{\mathrm{y}\prime} \right] = \sin ^2\frac{\zeta}{2}\frac{\Omega _{+}^{\prime}}{\frac{\omega _{\mathrm{m}}}{\cos \theta}}\left( 2n_++1 \right) +\cos ^2\frac{\zeta}{2}\frac{\Omega _{-}^{\prime}}{\frac{\omega _{\mathrm{m}}}{\cos \theta}}\left( 2n_-+1 \right) 
\end{equation}
\begin{equation}
\begin{aligned}
f_{\mathrm{Q}}\left[ |n_+,n_-\rangle ,S^{\mathrm{z}\prime} \right]
=&
\frac{4}{N_{\mathrm{B}}}\left( v_{21}u_{21} \right) ^2n_+\left( n_+-1 \right) +\left( v_{21}u_{21} \right) ^2\left( n_++2 \right) \left( n_++1 \right)
\\
&+
\frac{4}{N_{\mathrm{B}}}\left( u_{22}v_{22} \right) ^2\left( n_-+1 \right) \left( n_-+2 \right) +\left( u_{22}v_{22} \right) ^2n_-\left( n_--1 \right)
\\
&+
\frac{4}{N_{\mathrm{B}}}\left( u_{21}u_{22}+v_{22}v_{21} \right) ^2\left( n_++1 \right) n_-+\left( u_{22}u_{21}+v_{21}v_{22} \right) ^2n_+\left( n_-+1 \right)
\\
&+
\frac{4}{N_{\mathrm{B}}}\left( u_{21}v_{22}+u_{22}v_{21} \right) ^2n_+n_-+\left( u_{21}v_{22}+u_{22}v_{21} \right) ^2\left( n_++1 \right) \left( n_-+1 \right)
\end{aligned}
\end{equation}
We can similarly ignore $f_{\mathrm{Q}}\left[ |n_+,n_-\rangle ,S^{\mathrm{z}\prime} \right]$ ans thus 
\begin{align}
f_{\mathrm{Q}}\left[ |n_+,n_-\rangle ,S^{\mathrm{x}} \right] 
=&
\cos ^2\theta f_{\mathrm{Q}}\left[ |n_+,n_-\rangle ,S^{\mathrm{x}\prime} \right] 
\\
f_{\mathrm{Q}}\left[ |n_+,n_-\rangle ,S^{\mathrm{y}} \right] 
=&
f_{\mathrm{Q}}\left[ |n_+,n_-\rangle ,S^{\mathrm{y}\prime} \right] 
\\
f_{\mathrm{Q}}\left[ |n_+,n_-\rangle ,S^{\mathrm{z}} \right] 
=&
\sin ^2\theta f_{\mathrm{Q}}\left[ |n_+,n_-\rangle ,S^{\mathrm{x}\prime} \right] 
\end{align}
So QFI if linearly dependent on polariton occupation number $n_+$ and $n_-$.

\end{document}